\documentclass[11pt,onecolumn,draftclsnofoot]{IEEEtran}

\usepackage{amsfonts}
\usepackage{amssymb}
\usepackage{cite}
\usepackage{graphicx}
\usepackage[cmex10]{amsmath}
\usepackage[]{algorithm}
\usepackage{algorithmic}

\usepackage{soul}
\usepackage{changebar}
\usepackage{color}

\ifCLASSINFOpdf
\else
\fi

\begin{document}
%
\title{Projection Design For Statistical Compressive Sensing: A Tight Frame Based Approach}
%
%
%

\author{Wei~Chen,
        Miguel R. D.~Rodrigues,~\IEEEmembership{Member,~IEEE,}
        and~Ian J.~Wassell
\thanks{W. Chen and I. J. Wassell are with the Digital Technology Group, Computer Laboratory,
University of Cambridge, Cambridge CB3 0FD, United Kingdom (e-mail: {wc253, ijw24}@cam.ac.uk).}
\thanks{Miguel R. D. Rodrigues is with the Department of Electronic and Electrical Engineering, University College London,
London WC1E 7JE, United Kingdom (e-mail: m.rodrigues@ucl.ac.uk).}
\thanks{The work of M. R. D. Rodrigues was supported by Funda\c{c}\~{a}o para a Ci\^{e}ncia e a Tecnologia through the research
projects PTDC/EEA-TEL/100854/2008 and CMU-PT/RNQ/0029/2009.}

\thanks{A part of this work was presented at the 2012 IEEE International Conference on Acoustics,
Speech and Signal Processing (ICASSP), Kyoto, Japan, March 2012.}
}
\maketitle

\begin{abstract}
In this paper, we develop a framework to design sensing matrices for compressive sensing applications that lead to good
mean squared error (MSE) performance subject to sensing cost constraints. By capitalizing on the MSE of the oracle
estimator, whose performance has been shown to act as a benchmark to the performance of standard sparse recovery
algorithms, we use the fact that a Parseval tight frame is the closest design - in the Frobenius norm sense - to the
solution of a convex relaxation of the optimization problem that relates to the minimization of the MSE of the oracle
estimator with respect to the equivalent sensing matrix, subject to sensing energy constraints. Based on this result,
we then propose two sensing matrix designs that exhibit two key properties: i) the designs are closed form rather than
iterative; ii) the designs exhibit superior performance in relation to other designs in the literature, which is
revealed by our numerical investigation in various scenarios with different sparse recovery algorithms including basis
pursuit de-noise (BPDN), the Dantzig selector and orthogonal matching pursuit (OMP).
\end{abstract}



%
\IEEEpeerreviewmaketitle

\section{Introduction}
%
%
%
%

\IEEEPARstart{T}{he} presence of redundancy in most signals in nature offers the means to transform the original
signals into a compressed version convenient for storage and transportation. Compressive sensing (CS) is a new sampling
paradigm that, instead of conforming to the traditional two-stage process involving signal sampling followed by signal
compression, directly acquires a compressed version of the original signal instead, by leveraging signal sparsity (a
form of redundancy) as well as random sensing or measurement. In fact, it has been shown that if an $n$-dimensional
signal admits an $s$-sparse representation then one can reconstruct exactly the original signal with
$m=\mathcal{O}(s\log(n/s))$ measurements~\cite{1580791,1614066}; Also, if the original signal admits only a nearly
sparse representation (and/or the measurements are corrupted by some noise) then one can still reconstruct the original
signal subject to a tolerable distortion~\cite{1580791,1614066}. Therefore, CS offers the prospect of a more efficient
signal acquisition in relation to traditional Shannon-Nyquist sampling, especially in applications where the sampling
process is expensive such as magnetic resonance imaging~\cite{lustig2007sparse} and data acquisition in wireless sensor
networks~\cite{6287532}.

\par
A recent growing trend relates to the use of more complex signal models that go beyond the simple
sparsity model to further enhance the performance of CS. For example, Baraniuk et
al.~\cite{5437428} have introduced model-based compressive sensing, where more realistic signal
models such as wavelet trees or block sparsity are leveraged in order to reduce the number of
measurements required for reconstruction. In particular, it has been shown that robust signal
recovery is possible with $m=\mathcal{O}(s)$ measurements in model-based compressive
sensing~\cite{5437428}. Ji et al.~\cite{4524050} introduced Bayesian compressive sensing, where a
signal specific statistical model is exploited to reduce the number of measurements needed for
reconstruction. In~\cite{hegde2012signal,5559508}, reconstruction methods have been proposed for
manifold-based CS, where the signal is assumed to belong to a manifold. Other works that consider
various sparsity models that go beyond simple sparsity in order to improve the performance of
traditional CS
include~\cite{5771110,6094210,5887383,6155613,danielyan2008image,4379013,danielyan2010spatially}.

\par
The use of additional signal knowledge also enables one to replace the conventional random sensing matrices by
optimized ones in order to further enhance CS performance (e.g.,
see~\cite{4359525,5061489,xu2010optimized,5872076,carson2012communications}). A number of conditions have been put
forth to study the impact of the sensing matrices in various recovery algorithms. The null space property represents a
necessary and sufficient condition for sparse recovery~\cite{1614066}. However, it is difficult to verify whether or
not a certain sensing matrix fulfills this condition. Other more widely used conditions include the restricted isometry
property (RIP)~\cite{1580791}, which is also difficult to evaluate, and the mutual coherence~\cite{1564423}, which is
easier to evaluate. However, the fact that these conditions are mainly used to address the worst-case rather than the
expected-case performance, renders their use as the basis of sensing matrix designs as too conservative. As such,
Calderbank et al.~\cite{5419073} have put forth a weaker version of the RIP, the statistical restricted isometry
property (StRIP), where a probability criterion replaces the hard requirement demanded by RIP. StRIP has been also used
as the basis of various sensing matrix designs presented in~\cite{5419073}.

\par
In this paper, we develop a general framework to design sensing matrices for compressive sensing applications that lead
to good (expected-case) mean squared error (MSE) performance subject to sensing energy constraints, where the
expectation is with respect to both the statistical distribution of the signal and the noise. We also leverage
additional signal knowledge, by considering a general random signal model where the distinct support patterns of the
same sparsity level occur with equal probability in the sparse representation of the original signal, and the
autocorrelation matrix of the sparse representation is equal to an identity matrix. Our approach is based on the
analysis of the oracle estimator MSE~\cite{candes2007dantzig}, whose performance has been shown to act as a benchmark
to the performance of various common sparse recovery algorithms. By showing that good equivalent sensing matrices (that
correspond to the product of the sensing matrix and the sparsifying dictionary) ought to be close to a Parseval tight
frame, we are then able to put forth two new sensing matrix designs that conform to specific sensing energy
constraints. Our experiments reveal that the proposed designs improve signal expected-case reconstruction performance
in relation to random designs or other optimized designs~\cite{4359525,5061489,xu2010optimized}. Another notable
advantage of our proposed designs is that they are closed-form whereas the designs
in~\cite{4359525,5061489,xu2010optimized} are iterative.

\par
Our design approach, which is applicable to signals that are sparse in any dictionary, shares some of the elements of
the design approach in~\cite{6061944}, which is only applicable to signals that are sparse in an orthonormal basis. In
particular, this contribution - as does~\cite{6061944} - also capitalizes on the oracle estimator MSE to put forth
adequate sensing matrix designs. However, this design approach also departs significantly from that in~\cite{6061944},
in view of the fact that it is not clear how to generalize the methodology in~\cite{6061944} from orthonormal to
overcomplete dictionaries (namely, Propositions 1 and 2 in~\cite{6061944}).

\par
Therefore, the current generalization is based on two questions that are answered in the article.
We first ask:

\noindent 1) What is the equivalent sensing matrix that leads to the lowest oracle estimator MSE
for a certain target signal to noise ratio (SNR) at the input of the oracle estimator?

\noindent Further, in view of the fact that a Parseval tight frame is likely to provide a low oracle MSE subject to a
target SNR at the input of the oracle, we then ask:

\noindent 2) What is the sensing matrix that offers the best compromise between ``sensing cost"
and ``closeness" of the equivalent sensing matrix to a Parseval tight frame?

\par
It is this angle-of-attack - which departs from that in~\cite{6061944} - that enables us to
generalize the sensing matrix designs for signals that are sparse in arbitrary overcomplete
dictionaries. Interestingly, the ensuing designs are shown to reduce to the designs
in~\cite{6061944} when the dictionary is orthonormal rather than overcomplete.

\par
The generalization of the work from the orthonormal to overcomplete dictionary case is relevant not only theoretically
but also practically. For example, allowing signals to be sparse in overcomplete dictionaries adds a lot of flexibility
and extends the range of applicability for CS~\cite{rauhut2008compressed,Candes201159,davenport2012signal}. Of
particular relevance, this generalization also leads to further insight about the behavior of random vs. optimized
projections: this is also crisply exposed in this contribution.

\par
The rest of this paper is organized as follows. We begin by describing the CS model and assumptions in
Section~\uppercase\expandafter{\romannumeral2}. Section~\uppercase\expandafter{\romannumeral3} provides the rationale
for the sensing matrix designs, by highlighting the role of Parseval tight frames in compressive sensing applications.
Section~\uppercase\expandafter{\romannumeral4} puts forth our proposed sensing matrix designs, which capitalize on the
intuition unveiled in Section~\uppercase\expandafter{\romannumeral3}. Section~\uppercase\expandafter{\romannumeral5}
presents a range of numerical results that highlight the merits of our proposed designs in relation to other designs in
the literature. Section~\uppercase\expandafter{\romannumeral6} discusses the MSE performance yielded both by random and
optimized projections designs. The main contributions of the article are finally summarized in
Section~\uppercase\expandafter{\romannumeral7}.

\par
Throughout this paper, signals are treated as real-valued vectors. Lower-case letters denote
scalars, boldface upper-case letters denote matrices, bold face lower-case letters denote column
vectors, and calligraphic upper-case letters denote support sets. $\mathbf{0}$ and $\mathbf{1}$
denote a vector with all zeros and all ones, respectively, and $\mathbf{O}_{m \times n}$ denotes
an $m\times n$ matrix with all zeros. The superscripts $(\cdot)^T$ and $(\cdot)^{-1}$ denote
matrix transpose and matrix inverse, respectively. The $\ell_0$ norm, the $\ell_1$ norm, and the
$\ell_2$ norm of vectors, are denoted by $\|\cdot\|_{0}$, $\|\cdot\|_{1}$, and $\|\cdot\|_{2}$,
respectively. The Frobenius norm and spectral norm of a matrix $\mathbf{A}$ are denoted by
$\|\mathbf{A}\|_F$ and $\|\mathbf{A}\|$, respectively. The rank and trace of a matrix are denoted
by $\texttt{rank}(\cdot)$ and $\texttt{Tr}(\cdot)$, respectively. The diagonal matrix with
diagonal elements given by either vector $\mathbf{a}$ or the diagonal elements of matrix
$\mathbf{A}$ is denoted by $\text{Diag}(\mathbf{a})$ or $\text{Diag}(\mathbf{A})$, respectively.
The element corresponding to the $\emph{i}$th row and $\emph{j}$th column of the matrix
$\mathbf{A}$ is denoted by $a_{i,j}$, and $\mathbf{a}_i$ denotes the $\emph{i}$th column of the
matrix $\mathbf{A}$. $\mathbf{I}_n$ denotes the $n\times n$ identity matrix, and $\mathbf{J}_n$
denotes the $n\times n$ anti-diagonal matrix (an identity matrix with a reversed order of the
columns (or rows)). $\mathbf{E}_{\mathcal{J}}$ denotes the matrix that results from the identity
matrix by deleting the set of columns out of the support $\mathcal{J}$. $\mathbb{E}(\cdot)$
denotes the expectation, $\mathbb{E}_{\mathbf{x}}(\cdot)$ and $\mathbb{E}_{\mathcal{J}}(\cdot)$
denote expectation with respect to the distribution of the random vector $\mathbf{x}$, and the
random support $\mathcal{J}$, respectively. $\binom{n}{m}$ denotes the number of $m$ combinations
from a given set of $n$ elements. $\text{Pr}(\cdot)$ denotes the probability. Finally,
$\mathcal{N}(\boldsymbol{\mu},\boldsymbol{\Sigma})$ denotes the multivariate normal distribution
with mean vector $\boldsymbol{\mu}$ and covariance matrix $\boldsymbol{\Sigma}$.

%
%

\section{Compressive Sensing Model}
We consider the standard measurement model given by:
\begin{equation}\label{eq:CS_3_model}
\mathbf{y}=\mathbf{\Phi f}+\mathbf{n},
\end{equation}
where $\mathbf{y}\in \mathbb{R}^m$ is the measurement signal vector, $\mathbf{f}\in \mathbb{R}^n$ is the original
signal vector, $\mathbf{n}\sim\mathcal{N}(\mathbf{0},\sigma^2\mathbf{I}_m)\in \mathbb{R}^m$ is a zero-mean white
Gaussian noise vector, and $\mathbf{\Phi}\in\mathbb{R}^{m \times n}$ (with $m \leq n$) is the sensing matrix. We assume
that the original signal is sparse in some basis, i.e.,
\begin{equation}\label{eq:CS_3_signal_basis}
\mathbf{f}=\mathbf{\Psi x},
\end{equation}
where $\mathbf{\Psi}\in\mathbb{R}^{n \times \hat{n}}$ ($\hat{n}\geq n$) is a matrix that represents the sparsifying
basis, e.g., an orthonormal or overcomplete dictionary, and $\mathbf{x}\in \mathbb{R}^{\hat{n}}$ is a sparse
representation of $\mathbf{f}\in \mathbb{R}^n$, i.e., $\|\mathbf{x}\|_{0}\leq s\ll \hat{n}$. Then we can rewrite the
measurement model as
\begin{equation}\label{eq:CS_3_model2}
\mathbf{y}=\mathbf{\Phi \Psi x}+\mathbf{n}=\mathbf{A x}+\mathbf{n},
\end{equation}
where $\mathbf{A}=\mathbf{\Phi \Psi}\in\mathbb{R}^{m \times \hat{n}}$ represents the equivalent sensing matrix. For
modeling the sparse sources, we assume i) the distinct support patterns of the same sparsity level occur with equal
probability in the sparse representation of the original signal, i.e., $\text{Pr}\left(\mathcal{J}_c^t\right)=P_c$,
where $\mathcal{J}_c^t\subset\left\{1,\ldots,\hat{n}\right\}$ $(t=1,\ldots,\binom{\hat{n}}{c}$, $c=1,\ldots,s)$ denotes
a signal support with cardinality $c$ and $\sum_{c=1}^s\binom{\hat{n}}{c}P_c=1$; ii)
$\mathbb{E}_{\mathbf{x}}(\mathbf{x}\mathbf{x}^T)=\mathbf{I}_{\hat{n}}$. Note that these assumptions can be satisfied by
a signal model akin to the widely used Bernoulli-Gaussian
model\cite{553473,5072251,5398963,5484983,1542488,hyder2010improved,5930380}. In particular, one constrains the
cardinality of the support patterns to be less than $s$, rather than $\hat{n}$; one also constrains the probability of
the support patterns to obey $\sum_{c=1}^s\binom{\hat{n}}{c}P_c=1$ rather than a binomial distribution as in the
Bernoulli-Gaussian model.


\par
To recover the sparse signal representation $\mathbf{x}$ from the measurement vector $\mathbf{y}$, one can resort to
the optimization problem:
\begin{equation}\label{eq:CS_2_BPDN}
\begin{split}
\min_{\mathbf{x}} \qquad &\|\mathbf{x}\|_{1}\\
\text{s.t.} \qquad &\|\mathbf{A}\mathbf{x}-\mathbf{y}\|_{2}\leq \epsilon,
\end{split}
\end{equation}
where $\epsilon$ is an estimate of the noise level. This program is also known as the basis pursuit de-noise
(BPDN)~\cite{chen2001atomic}.

\par
It has been established in~\cite{candes2008restricted} that the now well-known RIP, which has been introduced by
Cand{\`e}s and Tao~\cite{1542412}, provides a sufficient condition for exact or near exact recovery of a sparse signal
representation $\mathbf{x}$ from the measurement vector $\mathbf{y}$ via the $\ell_1$ minimization in
(\ref{eq:CS_2_BPDN}).

\newtheorem{definition}{Definition}
\begin{definition}
A matrix $\mathbf{A}\in\mathbb{R}^{m\times \hat{n}}$ satisfies the RIP of order $s$ with a restricted isometry constant
(RIC) $\delta_s\in(0,1)$ being the smallest number such that
\begin{equation}\label{eq:CS_2_RIP}
(1-\delta_s)\|\mathbf{x}\|_{2}^2\leq\|\mathbf{Ax}\|_{2}^2\leq (1+\delta_s)\|\mathbf{x}\|_{2}^2
\end{equation}
holds for all $\mathbf{x}$ with $\|\mathbf{x}\|_{0}\leq s$.
\end{definition}

\newtheorem{theorem}{Theorem}
\begin{theorem}
The solution $\mathbf{x}^*$ of (\ref{eq:CS_2_BPDN}) obeys
\begin{equation}\label{eq:CS_2_sta3}
\|\mathbf{x}^*-\mathbf{x}\|_{2} \leq c_1s^{-1/2}\|\mathbf{x}-\mathbf{x}_s\|_{1}+c_2\epsilon,
\end{equation}
where $c_1=\frac{2+(2\sqrt{2}-2)\delta_{2s}}{1-(\sqrt{2}+1)\delta_{2s}}$,
$c_2=\frac{4\sqrt{1+\delta_{2s}}}{1-(\sqrt{2}+1)\delta_{2s}}$, $\mathbf{x}_s$ is an approximation of $\mathbf{x}$ with
all but the $s$-largest entries set to zero, and $\delta_{2s}$ is the RIC of order $2s$ of matrix $\mathbf{A}$.
\end{theorem}

\par
This theorem claims that the reconstructed signal representation $\mathbf{x}^*$ is a good
approximation to the original signal representation $\mathbf{x}$. In addition, for the noiseless
case, any sparse representation $\mathbf{x}$ with support size no larger than $s$, can be exactly
recovered by $\ell_1$ minimization if the RIC satisfies $\delta_{2s}<\sqrt{2}-1$. Therefore, it
follows that the RIP acts as a proxy to the quality of a sensing matrix. Note that the RIP is a
sufficient condition for successful reconstruction but it may be too strict. It has been observed
that signals with sparse representations can be reconstructed very well even though the sensing
matrices have not been proven to satisfy the RIP~\cite{5419073}.

\par
Another way to evaluate a sensing matrix, which is not as computationally intractable as the RIP,
is via the mutual coherence of the matrix $\mathbf{A}$, given by~\cite{1564423}:
\begin{equation}\label{eq:CS_2_MC}
\mu=\max_{1\leq i,j\leq \hat{n},i\neq j}|\mathbf{a}_i^T\mathbf{a}_j|.
\end{equation}
Donoho, Elad and Temlyakov~\cite{1564423} demonstrated that the error of the solution to
(\ref{eq:CS_2_BPDN}) is bounded if $\mu<\frac{1}{4s-1}$. Therefore, mutual coherence can also be
used to measure the quality of a sensing matrix. For example, various sensing matrix design
approaches in the literature, such as Elad's method~\cite{4359525}, Duarte-Carvajalino and
Sapiro's method ~\cite{5061489}, and Xu et al.'s method~\cite{xu2010optimized} are inherently
mutual coherence based approaches.


\section{Design Rationale}
We now provide a rationale for the proposed novel sensing matrix designs. The ultimate goal of the sensing matrix
designs relates to the minimization of the MSE in estimating $\mathbf{x}$ from $\mathbf{y}$, given by
\begin{equation}\label{eq:CS_4_MSE_signal_noise}
\texttt{MSE}(\mathbf{\Phi})=\mathbb{E}_{\mathbf{x},\mathbf{n}}\left(\|\mathcal{F}(\mathbf{\Phi}\mathbf{\Psi}\mathbf{x}+\mathbf{n})-\mathbf{x}\|_{2}^2\right),
\end{equation}
where $\mathcal{F}(\cdot)$ denotes an estimator, subject to appropriate constraints (e.g., sensing energy
cost)\footnote{We would also like to add that one could argue that it is preferable to consider the MSE associated with
the estimation of $\mathbf{f}$ (the actual signal) from $\mathbf{y}$ rather than the MSE associated with the estimation
of $\mathbf{x}$ (the signal sparse representation) from $\mathbf{y}$. We use the more tractable MSE associated with the
estimation of $\mathbf{x}$ from $\mathbf{y}$ because: 1) it can be shown that the MSE performance associated with the
(oracle) estimation of $\mathbf{x}$ from $\mathbf{y}$ upper bounds in general the MSE performance associated with the
(oracle) estimation of $\mathbf{f}$ from $\mathbf{y}$. In particular, for an orthogonal dictionary, where
$\mathbf{\Psi}$ is an orthogonal matrix, $||\mathbf{f}-\mathbf{f}^*||_2^2=||\mathbf{\Psi x}-\mathbf{\Psi
x}^*||_2^2=||\mathbf{x}-\mathbf{x}^*||_2^2$, where $\mathbf{x}^*$ denotes the (oracle) estimate of $\mathbf{x}$ and
$\mathbf{f}^*=\mathbf{\Psi x}^*$ denotes the (oracle) estimate of $\mathbf{f}$; for an overcomplete dictionary, where
$\mathbf{\Psi}$ is not an orthogonal matrix, $||\mathbf{f}-\mathbf{f}^*||_2^2=||\mathbf{\Psi x}-\mathbf{\Psi
x}^*||_2^2\leq \lambda_{\max}^2(\mathbf{\Psi}) ||\mathbf{x}-\mathbf{x}^*||_2^2 $, where $\lambda_{\max}(\mathbf{\Psi})$
is the largest singular value of $\mathbf{\Psi}$; 2) it is also often desirable to manipulate or process the
information content of signals in the sparse representation domain rather than the original observation domain, such as
in feature extraction, pattern classification and blind source separation~\cite{4801665,1300802,847906}. Therefore, the
MSE performance associated with the estimation of $\mathbf{x}$ would be more appropriate than the MSE performance
associated with the estimation of $\mathbf{f}$ for such applications.}.

\par
The derivation of such a sensing matrix design is very difficult though, because the average MSE in
(\ref{eq:CS_4_MSE_signal_noise}) depends upon the actual estimator. Consequently, to avoid the analysis of a single or
several practical sparse recovery algorithms such as the BPDN, the Dantzig selector, or the OMP, we capitalize - as
in~\cite{6061944} - on the well-known oracle estimator that performs ideal least squares (LS) estimation based on prior
knowledge of the sparse vector support $\mathcal{J}\subset\{1,\ldots,\hat{n}\}$~\cite{candes2007dantzig}. The rationale
of this approach is supported by the fact that the MSE of this oracle LS estimator coincides with the unbiased
Cram{\'e}r-Rao bound (CBD) for exactly $s$-sparse deterministic vectors~\cite{5428818}, so that it represents the best
achievable performance for any unbiased estimator. Equally important, this approach is also supported by the fact that
the oracle estimator MSE performance acts as a performance benchmark for the key sparse recovery algorithms. For
example, Ben-Haim, Eldar and Elad~\cite{5483095} demonstrate both theoretically and numerically that the BPDN, the
Dantzig selector, the OMP and thresholding algorithms all achieve performances that are proportional to the oracle
estimator MSE.

\par
The oracle estimator MSE incurred in the estimation of a sparse deterministic vector $\mathbf{x}$
in the presence of a standard Gaussian noise vector $\mathbf{n}$, according to the model in
(\ref{eq:CS_3_model}), is given by~\cite{candes2007dantzig}\footnote{Note that various works have
adopted the oracle minimum MSE (MMSE) estimator in lieu of the oracle LS one in order to obtain a
superior MMSE estimate~\cite{4063557,schniter2009fast,5238753}. The fact that we assume a signal
model that does not specify the exact distribution of the sparse signal conditioned on the support
- in contrast to~\cite{4063557,schniter2009fast,5238753} that take the distribution of the sparse
signal conditioned on the support to be multi-variate Gaussian - prevents us from exploiting this
more powerful estimator. This approach however instils our projections design framework with more
generality.}:
\begin{equation}\label{eq:CS_MSE_oracle}
\begin{split}
\texttt{MSE}^{\text{oracle}}_{\mathbf{n}}(\mathbf{A},\mathbf{x})=&\mathbb{E}_{\mathbf{n}}\left(\|\mathcal{F}^{\text{oracle}}(\mathbf{A}\mathbf{x}+\mathbf{n})-\mathbf{x}\|_{2}^2\right)\\
=&\sigma^2\text{Tr}\left(\left(\mathbf{E}_{\mathcal{J}}^T\mathbf{A}^T\mathbf{A}\mathbf{E}_{\mathcal{J}}\right)^{-1}\right).
\end{split}
\end{equation}
Consequently, the average value of the oracle estimator MSE incurred in the estimation of a sparse random vector
$\mathbf{x}$ in the presence of the Gaussian noise vector $\mathbf{n}$ is given by:
\begin{equation}\label{eq:CS_4_MSE_oricle_1}
\texttt{MSE}^{\text{oracle}}(\mathbf{A})=\sigma^2\mathbb{E}_{\mathcal{J}}\left(\text{Tr}\left(\left(\mathbf{E}_{\mathcal{J}}^T
\mathbf{A}^T\mathbf{A}\mathbf{E}_{\mathcal{J}}\right)^{-1}\right)\right).
\end{equation}

\par
We define the coherence matrix of the equivalent sensing matrix as
$\mathbf{Q}=\mathbf{A}^T\mathbf{A}=\mathbf{\Psi}^T\mathbf{\Phi}^T\mathbf{\Phi}\mathbf{\Psi}$. We now pose the
optimization problem:
\begin{equation}\label{eq:CS_4_optMSE_oracle}
\begin{split}
\min_{\mathbf{Q}}& \quad \mathbb{E}_{\mathcal{J}}\left(\text{Tr}\left(\left(\mathbf{E}_{\mathcal{J}}^T
\mathbf{Q}\mathbf{E}_{\mathcal{J}}\right)^{-1}\right)\right)\\
\text{s.t.}& \quad \mathbf{Q}\succeq 0,\\
& \quad \text{Tr}\left(\mathbf{Q}\right)=m, \\
& \quad rank(\mathbf{Q})\leq m.
\end{split}
\end{equation}

\par
It is relevant to reflect further on the rationale of this optimization problem. This optimization problem defines the
coherence matrix of the equivalent sensing matrix - up to a rotation - that minimizes the average value of the oracle
MSE subject to appropriate constraints: these include the obvious positive semi-definite and rank constraints on the
coherence matrix and - at the heart of the novelty of the approach - a trace constraint on the coherence matrix that
acts as a proxy to the sensed energy.

\par
In the noiseless case~\cite{4359525,5061489,xu2010optimized}, it is not common to place a constraint on the sensed
energy because recovery is immune to the scaling of the sensing matrix; instead, it is only common to seek sensing
matrices that exhibit adequate structure (e.g., \cite{4359525} uses $t$-averaged mutual coherence, \cite{5061489} uses
an equivalent sensing matrix whose Gram matrix is similar to an identity matrix, and \cite{xu2010optimized} uses an
equivalent sensing matrix which is close to an equiangular tight frame, to seek for sensing matrices with adequate
structure).

\par
In contrast, in the noisy case it is important to place a constraint on the sensed energy because recovery is
affected both by the sensing matrix structure and immunity to noise. Therefore, the main features of our formulation
include:
\begin{enumerate}
  \item The optimization problem defines equivalent sensing matrices with good structure and immunity to noise.
  \item The formulation is such that the sensed energy is directly proportional to the number of measurements. In fact, the
sensed energy is given by:
\begin{equation}\label{eq:CS_4_optMSE_eq1}
\begin{split}
\mathbb{E}_{\mathbf{x}}\left(\text{Tr}\left(\mathbf{\Phi\Psi
x}\mathbf{x}^T\mathbf{\Psi}^T\mathbf{\Phi}^T\right)\right)&=\text{Tr}\left(\mathbf{\Phi\Psi}\mathbb{E}_{\mathbf{x}}\left(\mathbf{x
x}^T\right)\mathbf{\Psi}^T\mathbf{\Phi}^T\right)\\
&=\text{Tr}\left(\mathbf{\Phi\Psi}\mathbf{\Psi}^T\mathbf{\Phi}^T\right)=m,
\end{split}
\end{equation}
where we have used the fact that $\mathbb{E}_{\mathbf{x}}\left(\mathbf{x x}^T\right)=\mathbf{I}_{\hat{n}}$. Note that a
modification of the constant of proportionality, which is equal to 1 here, scales only the solution to the optimization
problem (\ref{eq:CS_4_optMSE_oracle}).
  \item The formulation is also such that the sensed SNR
\begin{equation}\label{eq:CS_4_optMSE_snr}
\frac{\mathbb{E}_{\mathbf{x}}\left(\text{Tr}\left(\mathbf{\Phi\Psi
x}\mathbf{x}^T\mathbf{\Psi}^T\mathbf{\Phi}^T\right)\right)}{\mathbb{E}_{\mathbf{n}}\left(\text{Tr}\left(\mathbf{n}\mathbf{n}^T\right)\right)}
=\frac{1}{\sigma^2},
\end{equation}
does not depend on $m$, $n$ or $\hat{n}$.
\end{enumerate}
We will see that in the presence of noise some of the ``noiseless" sensing matrix designs in the literature can yield
very poor recovery performance (see Section V). This is due to the fact that upon the normalization of the sensing
matrix so that it conforms to a specific sensing cost constraint, the structural properties of the designs are offset
by the poor noise immunity of the designs. The optimization problem formulation in (\ref{eq:CS_4_optMSE_oracle}) aims
thus to attain a compromise between the structural and the noise immunity properties of the sensing
matrix\footnote{Note that this optimization problem places a cost on the equivalent sensing matrix
$\mathbf{A}=\mathbf{\Phi\Psi}$, which translates into a constraint on the energy given to the estimator rather than a
cost on the sensing matrix $\mathbf{\Phi}$, which translates into a constraint on the sensing energy. We recognize that
a sensing energy cost is often more appropriate, but this is difficult to analyze in general. Therefore, our approach
when the signal is sparse in a general overcomplete dictionary departs from that when the signal is sparse in an
orthonormal dictionary~\cite{6061944}. In particular, we only incorporate the effect of sensing energy constraints into
the design framework in Section~\uppercase\expandafter{\romannumeral4}.}.

\par
The optimization problem (\ref{eq:CS_4_optMSE_oracle}) is non-convex owing to the rank constraint, and so is very
difficult to solve. Therefore, we adopt an approach akin to that in~\cite{6061944}: i) we first consider a convex
relaxation of (\ref{eq:CS_4_optMSE_oracle}) by ignoring the rank constraint; and ii) we then consider the feasible
solution that is closest to the solution to the relaxed problem. This procedure produces a sub-optimal equivalent
sensing matrix, but extensive simulation results demonstrate that this design outperforms various other designs.

\newtheorem{proposition}{Proposition}
\begin{proposition}\label{thm:CS_4_optimization}
The solution of the optimization problem:
\begin{equation}\label{eq:CS_4_optMSE_Proposition}
\begin{split}
\min_{\mathbf{Q}}& \quad
\mathbb{E}_{\mathcal{J}}\left(\text{Tr}\left(\left(\mathbf{E}_{\mathcal{J}}^T
\mathbf{Q}\mathbf{E}_{\mathcal{J}}\right)^{-1}\right)\right)\\
\text{s.t.}& \quad \mathbf{Q}\succeq 0,\\
& \quad \text{Tr}\left(\mathbf{Q}\right)=m,
\end{split}
\end{equation}
which represents a convex relaxation of the original optimization problem in (\ref{eq:CS_4_optMSE_oracle}), is the
$\hat{n}\times \hat{n}$ matrix $\frac{m}{\hat{n}}\mathbf{I}_{\hat{n}}$.
\end{proposition}

\begin{IEEEproof}
See Appendix A.
\end{IEEEproof}

\par
It is evident that the solution to the convex relaxation of the original optimization problem is not feasible, because
$\texttt{rank}(\frac{m}{\hat{n}}\mathbf{I}_{\hat{n}})=\hat{n}\geq m$. Therefore, we now propose to determine the
$m\times \hat{n}$ matrix $\mathbf{A}$ whose $\hat{n}\times \hat{n}$ coherence matrix
$\mathbf{Q}=\mathbf{A}^T\mathbf{A}$ is closest to the $\hat{n}\times \hat{n}$ matrix
$\frac{m}{\hat{n}}\mathbf{I}_{\hat{n}}$.

\begin{proposition}\label{thm:CS_4_solution}
The solution of the optimization problem:
\begin{equation}\label{eq:CS_4_optmatrix_Proposition}
\begin{split}
\min_{\mathbf{A}}& \quad \left\|\mathbf{A}^T\mathbf{A}-\frac{m}{\hat{n}}\mathbf{I}_{\hat{n}}\right\|_F^2\\
\text{s.t.}& \quad \text{Tr}\left(\mathbf{A}^T\mathbf{A}\right)=m,
\end{split}
\end{equation}
is the $m\times \hat{n}$ Parseval tight frame.
\end{proposition}

\begin{IEEEproof}
See Appendix B.
\end{IEEEproof}

\par
A frame in a finite-dimensional real space can be seen as a matrix $\mathbf{A}\in
 \mathbb{R}^{m\times \hat{n}}$ such that for any vector $\mathbf{z}\in \mathbb{R}^m$,
\begin{equation}\label{eq:CS_3_Parseval_frame_1}
a\|\mathbf{z}\|_2^2\leq\|\mathbf{A}^T\mathbf{z}\|_2^2\leq b\|\mathbf{z}\|_2^2,
\end{equation}
where $a>0$ and $b>0$ are known as the frame bounds. Tight frames are a class of frames with equal frame bounds, i.e.,
$a=b$. A tight frame whose columns have unit $\ell_2$ norm is called a unit norm tight frame. A tight frame whose frame
bound is equal to 1, is called a Parseval tight frame. Note that any tight frame can be scaled by multiplying by
$\frac{1}{\sqrt{a}}$, so that the frame bound becomes equal to 1.

\par
Therefore, the value of the constraint of (\ref{eq:CS_4_optmatrix_Proposition}) leads to a frame with a frame bound
being equal to 1, and thus results in a Parseval tight frame. By scaling the value of
$\text{Tr}\left(\mathbf{A}^T\mathbf{A}\right)$ in the constraint, which in fact alters the target sensing SNR in
(\ref{eq:CS_4_optMSE_snr}), it is clear that the solution of the optimization problem
(\ref{eq:CS_4_optmatrix_Proposition}) is still a tight frame. Therefore, we can deduce that the tight frame represents
a good equivalent sensing matrix design, in the sense that, among all equivalent sensing matrices that conform to the
target sensing SNR, a tight frame is likely to produce a good MSE performance. Appendix C explores another facet of
tight frames, including the relationship of a unit-norm tight frame to StRIP.

\par
Note that an alternative way to prove Proposition~\ref{thm:CS_4_solution}, which has been motivated by the optimization
problem put forth by Duarte-Carvajalino and Sapiro~\cite{5061489}, is also provided in~\cite{5872076}. The current
problem differs from the problems in~\cite{5061489,5872076} since our optimization approach is based on a metric with
operational significance, the MSE, whereas the optimization approach in~\cite{5061489,5872076} is based on mutual
coherence.

\section{Novel Sensing Matrix Design Approaches}
We now build upon the previous analysis, which suggests that $\mathbf{A}=\mathbf{\Phi\Psi}$ ought
to be close to a Parseval tight frame, to propose two sensing matrix designs for the compressive
sensing model in (\ref{eq:CS_3_model2}). In particular, in view of the fact that it is usual to
place a constraint on the sensing energy cost $\|\mathbf{\Phi}\|_F^2=n$, the design approaches
strike a balance between two objectives: i) guaranteeing that the equivalent sensing matrix
$\mathbf{A}=\mathbf{\Phi\Psi}$ is as close as possible to a Parseval tight frame; and ii)
guaranteeing that the sensing cost $\|\mathbf{\Phi}\|_F^2$ is as small as possible. For example,
for two different sensing matrices $\mathbf{\Phi}'$ and $\mathbf{\Phi}''$ such that
$\mathbf{\Phi}'\mathbf{\Psi}=\mathbf{\Phi}''\mathbf{\Psi}$ is equal or close (e.g., in Frobenius
norm sense) to some Parseval tight frame and $\|\mathbf{\Phi}'\|_F^2<\|\mathbf{\Phi}''\|_F^2$, it
may be preferable to use $\mathbf{\Phi}'$ instead of $\mathbf{\Phi}''$ in the compressive sensing
model in (\ref{eq:CS_3_model2}). In fact, the normalization
\begin{equation}\label{eq:CS_4_normailization0}
\tilde{\mathbf{\Phi}}'=\frac{\sqrt{n}\mathbf{\Phi}'}{\|\mathbf{\Phi}'\|_F}
\end{equation}
and
\begin{equation}\label{eq:CS_4_normailization1}
\tilde{\mathbf{\Phi}}''=\frac{\sqrt{n}\mathbf{\Phi}''}{\|\mathbf{\Phi}''\|_F}
\end{equation}
then ensures that
\begin{equation}\label{eq:CS_4_normailization2}
\|\tilde{\mathbf{\Phi}}'\mathbf{\Psi}\|_F^2>\|\tilde{\mathbf{\Phi}}''\mathbf{\Psi}\|_F^2
\end{equation}
and - via the previous analysis - eventually
\begin{equation}\label{eq:CS_4_normailization3}
\texttt{MSE}^{\text{oracle}}(\tilde{\mathbf{\Phi}}'\mathbf{\Psi})<\texttt{MSE}^{\text{oracle}}(\tilde{\mathbf{\Phi}}''\mathbf{\Psi}).
\end{equation}

\par
We note that this design approach, which is applicable to the noisy setting, is fundamentally different from the
approaches in~\cite{4359525,5061489,xu2010optimized}, which in contrast apply to the noiseless case. In particular, our
design considers the sensing energy cost whereas the designs in~\cite{4359525,5061489,xu2010optimized} do not. We will
reveal the effect of taking into account the sensing energy constraint when we re-normalize the designs
in~\cite{4359525,5061489,xu2010optimized}, by showing the radically different performances in the presence of noise.

\subsection{Design Approach 1}
We now consider the first sensing matrix design approach, which explicitly performs a balance between the objective of
guaranteeing that the equivalent sensing matrix is as close as possible to a Parseval tight frame against the objective
of guaranteeing that the sensing energy cost is as small as possible. In particular, we pose the design problem:
\begin{equation}\label{eq:CS_4_optimization_overcomplete}
\min_{\hat{\mathbf{\Phi}}} \quad
\left\|\hat{\mathbf{\Phi}}\mathbf{\Psi}-\mathbf{B}\right\|_F^2+\alpha\left\|\hat{\mathbf{\Phi}}\right\|_F^2,
\end{equation}
where $\mathbf{B}\in \mathbb{R}^{m\times \hat{n}}$ is a specific target Parseval tight frame and $\alpha\geq0$ is a
specific scalar. The solution to the design problem is:
\begin{equation}\label{eq:CS_4_optimization_overcomplete2}
\hat{\mathbf{\Phi}}=\mathbf{B}\mathbf{\Psi}^T\left(\mathbf{\Psi}\mathbf{\Psi}^T+\alpha\mathbf{I}_n\right)^{-1}.
\end{equation}
In turn, the sensing matrix design, which is consistent with the sensing cost constraint $\|\mathbf{\Phi}\|_F^2=n$, is:
\begin{equation}\label{eq:CS_4_optimization_overcomplete3}
\mathbf{\Phi}=\frac{\sqrt{n}\hat{\mathbf{\Phi}}}{\|\hat{\mathbf{\Phi}}\|_F}
=\frac{\sqrt{n}\mathbf{B}\mathbf{\Psi}^T\left(\mathbf{\Psi}\mathbf{\Psi}^T+\alpha\mathbf{I}_n\right)^{-1}}
{\|\mathbf{B}\mathbf{\Psi}^T\left(\mathbf{\Psi}\mathbf{\Psi}^T+\alpha\mathbf{I}_n\right)^{-1}\|_F}.
\end{equation}
We note that the scalar $\alpha$ controls the weight for the energy penalty of the sensing matrix. If the penalty is
not considered, i.e., $\alpha=0$, we have the sensing matrix design
$\mathbf{\Phi}=\frac{\sqrt{n}\mathbf{B}\mathbf{\Psi}^T\left(\mathbf{\Psi}\mathbf{\Psi}^T\right)^{-1}}
{\|\mathbf{B}\mathbf{\Psi}^T\left(\mathbf{\Psi}\mathbf{\Psi}^T\right)^{-1}\|_F}$. In contrast, for a very high penalty,
i.e., $\alpha\rightarrow +\infty$, we have the design $\mathbf{\Phi}=\frac{\sqrt{n}\mathbf{B}\mathbf{\Psi}^T}
{\|\mathbf{B}\mathbf{\Psi}^T\|_F}$. In both cases, i.e., $\alpha=0$ or $\alpha\rightarrow +\infty$, the sensing matrix
$\mathbf{\Phi}$ turns out to be a unit norm tight frame if the basis $\mathbf{\Psi}$ is an orthonormal matrix and the
design target $\mathbf{B}$ is a tight frame with equal column norm, i.e., a scaled unit norm tight frame. We also note
that, as will be shown later, the performance gain is greatly affected by the parameter $\alpha$. In particular, one
needs to use some empirical knowledge in order to set a suitable value for $\alpha$. We next propose a sensing matrix
design approach, that does not contain any adjustable parameters.

\subsection{Design Approach 2}
We now consider the second sensing matrix design approach, where the objective is to determine the matrix design with
the lowest sensing energy cost that is consistent with the fact that the equivalent sensing matrix ought to be a
Parseval tight frame. It will be shown that the ensuing design is instilled with operational significance, akin to the
design in~\cite{carson2012communications}. We pose the design problem:
\begin{equation}\label{eq:CS_4_optimization_approach2}
\begin{split}
\min_{\hat{\mathbf{\Phi}}}& \quad \left\|\hat{\mathbf{\Phi}}\right\|_F^2\\
\text{s.t.}& \quad \hat{\mathbf{\Phi}}\mathbf{\Psi}\mathbf{\Psi}^T\hat{\mathbf{\Phi}}^T=\mathbf{I}_m.
\end{split}
\end{equation}

\par
The following Proposition defines the solution to this optimization problem. We use the singular value decomposition
(SVD) of the dictionary $\mathbf{\Psi}=\mathbf{U}_{\Psi}\mathbf{\Lambda}_{\Psi}\mathbf{V}_{\Psi}^T$, where
$\mathbf{U}_{\Psi}\in \mathbb{R}^{n\times n}$ and $\mathbf{V}_{\Psi}\in \mathbb{R}^{\hat{n}\times \hat{n}}$ are
orthonormal matrices, and $\mathbf{\Lambda}_{\Psi}\in \mathbb{R}^{n\times \hat{n}}$ is a matrix whose main diagonal
entries ($\lambda^{\Psi}_1\geq \lambda^{\Psi}_2\geq \ldots \lambda^{\Psi}_n\geq 0$) are the singular values of
$\mathbf{\Psi}$ and the other entries are zeros. We also use the SVD of the sensing matrix
$\hat{\mathbf{\Phi}}=\mathbf{U}_{\hat{\Phi}}\mathbf{\Lambda}_{\hat{\Phi}}\mathbf{V}_{\hat{\Phi}}^T$, where
$\mathbf{U}_{\hat{\Phi}}\in \mathbb{R}^{m\times m}$ and $\mathbf{V}_{\hat{\Phi}}\in \mathbb{R}^{n\times n}$ are
orthonormal matrices, and $\mathbf{\Lambda}_{\hat{\Phi}}\in \mathbb{R}^{m\times n}$ is a matrix whose main diagonal
entries ($\lambda^{\hat{\Phi}}_1\geq \lambda^{\hat{\Phi}}_2\geq \ldots \geq \lambda^{\hat{\Phi}}_m\geq 0$) are the
singular values of $\hat{\mathbf{\Phi}}$ and the other entries are zeros.

\begin{proposition}\label{thm:CS_4_approach2}
A sensing matrix that solves the optimization problem in (\ref{eq:CS_4_optimization_approach2}) is given by
\begin{equation}\label{eq:CS_4_approach2_design}
\hat{\mathbf{\Phi}}=\mathbf{U}_{\hat{\Phi}}\mathbf{\Lambda}_{\hat{\Phi}}\mathbf{J}_n\mathbf{U}_{\Psi}^T,
\end{equation}
where $\mathbf{U}_{\hat{\Phi}}$ is an arbitrary orthonormal matrix and
$\mathbf{\Lambda}_{\hat{\Phi}}=\left[\text{Diag}\left(\frac{1}{\lambda^{\Psi}_m},\ldots,\frac{1}{\lambda^{\Psi}_1}\right)\
\mathbf{O}_{m\times (n-m)}\right]$.
\end{proposition}

\begin{IEEEproof}
Consider the SVD of the dictionary $\mathbf{\Psi}=\mathbf{U}_{\Psi}\mathbf{\Lambda}_{\Psi}\mathbf{V}_{\Psi}^T$ and the
sensing matrix $\hat{\mathbf{\Phi}}=\mathbf{U}_{\hat{\Phi}}\mathbf{\Lambda}_{\hat{\Phi}}\mathbf{V}_{\hat{\Phi}}^T$.
Then the equivalent sensing matrix can be expressed as:
\begin{equation}\label{eq:CS_4_matrix_permutation}
\hat{\mathbf{\Phi}}\mathbf{\Psi}=\mathbf{U}_{\hat{\Phi}}\mathbf{\Lambda}_{\hat{\Phi}}\mathbf{V}_{\hat{\Phi}}^T\mathbf{U}_{\Psi}\mathbf{\Lambda}_{\Psi}\mathbf{V}_{\Psi}^T,
\end{equation}
and so the Parseval tight frame constraint in (\ref{eq:CS_4_optimization_approach2}) can also be expressed as:
\begin{equation}\label{eq:CS_4_approach2_constriant}
\mathbf{\Lambda}_{\hat{\Phi}}\mathbf{V}_{\hat{\Phi}}^T\mathbf{U}_{\Psi}\mathbf{\Lambda}_{\Psi}\mathbf{\Lambda}_{\Psi}^T\mathbf{U}_{\Psi}^T
\mathbf{V}_{\hat{\Phi}}\mathbf{\Lambda}_{\hat{\Phi}}^T=\mathbf{I}_m.
\end{equation}

\par
To satisfy the Parseval tight frame condition in (\ref{eq:CS_4_approach2_constriant}), it is clear that $m$ columns of
$\mathbf{V}_{\hat{\Phi}}$ have to correspond to $m$ columns of $\mathbf{U}_{\Psi}$. Since the remaining $n-m$ columns
of $\mathbf{V}_{\hat{\Phi}}$ do not affect the Parseval tight frame condition at all, then we take without any loss of
generality $\mathbf{V}_{\hat{\Phi}}=\mathbf{U}_{\Psi}\mathbf{\Pi}$, where $\mathbf{\Pi}\in \mathbb{R}^{n\times n}$ is a
permutation matrix. Therefore, we can now rewrite the optimization problem as follows:
\begin{equation}\label{eq:CS_4_optmatrix_approach2_new}
\begin{split}
\min_{\mathbf{\Lambda}_{\hat{\Phi}},\mathbf{\Pi}}& \quad \left\|\mathbf{\Lambda}_{\hat{\Phi}}\right\|_F^2\\
\text{s.t.}& \quad \mathbf{\Lambda}_{\hat{\Phi}}\mathbf{\Pi}^T\mathbf{\Lambda}_{\Psi}
\mathbf{\Lambda}_{\Psi}^T\mathbf{\Pi}\mathbf{\Lambda}_{\hat{\Phi}}^T=\mathbf{I}_m,\\
& \quad \mathbf{\Pi}\ \text{is a permutation matrix}.
\end{split}
\end{equation}

\par
The solution to this optimization problem is trivially given by:
\begin{equation}\label{eq:CS_4_approach2_Pf1}
\mathbf{\Pi}=\mathbf{J}_n,
\end{equation}
and
\begin{equation}\label{eq:CS_4_approach2_Pf2}
\begin{split}
\mathbf{\Lambda}_{\hat{\Phi}}=&\left[\text{Diag}\left(\lambda^{\hat{\Phi}}_1,\lambda^{\hat{\Phi}}_2, \ldots ,
\lambda^{\hat{\Phi}}_m\right)\ \mathbf{O}_{m\times (n-m)}\right]\\
=&\left[\text{Diag}\left(\frac{1}{\lambda^{\Psi}_m},\frac{1}{\lambda^{\Psi}_{m-1}}, \ldots
,\frac{1}{\lambda^{\Psi}_1}\right)\ \mathbf{O}_{m\times (n-m)}\right].
\end{split}
\end{equation}
\end{IEEEproof}

\par
Proposition~\ref{thm:CS_4_approach2} uncovers the key operations performed by this sensing matrix design. In
particular, this sensing matrix design i) exposes the modes (singular values) of the dictionary; ii) passes through the
$m$ strongest modes and filters out the $n-m$ weakest modes; and iii) weighs the strongest modes. This is accomplished
by taking the matrix of right singular vectors of the sensing matrix to correspond to the matrix of left singular
vectors of the dictionary and taking the strongest modes of the dictionary.

\par
Proposition~\ref{thm:CS_4_approach2} leads immediately to the sensing matrix design, which is consistent with the
sensing cost constraint $\|\hat{\mathbf{\Phi}}\|_F^2=n$, as follows:
\begin{equation}\label{eq:CS_4_approach2_normalized}
\mathbf{\Phi}=\frac{\sqrt{n}\hat{\mathbf{\Phi}}}{\|\hat{\mathbf{\Phi}}\|_F}
=\frac{\sqrt{n}\mathbf{\Lambda}_{\hat{\Phi}}\mathbf{J}_n\mathbf{U}_{\Psi}^T} {\|\mathbf{\Lambda}_{\hat{\Phi}}\|_F}.
\end{equation}

\par
Note that the design approach 1 balances the requirements of guaranteeing that the equivalent sensing matrix is as
close as possible to a Parseval tight frame against the requirements of guaranteeing that the cost is as small as
possible; in the design approach 2, we force the equivalent sensing matrix to be a Parseval tight frame and minimize
the sensing energy. Note also that the proposed designs are closed-form whereas other designs in the literature, such
as Elad's method~\cite{4359525}, Duarte-Carvajalino and Sapiro's method ~\cite{5061489}, and Xu et al.'s
method~\cite{xu2010optimized}, are iterative.

\par
Finally, it is also interesting to note that design 1 and design 2 reduce to the design in~\cite{6061944}, i.e., to a
tight frame, when we take the dictionary to be orthonormal rather than overcomplete.

\section{Performance Results}
We now compare the performance of the proposed sensing matrix designs to other designs in the CS
setting.

\subsection{Distribution of the off-diagonal entries of the coherence matrix}
We first investigate the histogram of the absolute values of the off-diagonal entries of the
coherence matrix $\mathbf{\Psi}^T\mathbf{\Phi}^T\mathbf{\Phi}\mathbf{\Psi}$. In this
investigation, we use a random dictionary $\mathbf{\Psi}\in\mathbb{R}^{64 \times 80}$ with entries
drawn from i.i.d. zero mean and unit variance Gaussian distributions and then normalized to
$\|\mathbf{\Psi}\|_F^2=80$. We also generate three sensing matrices
$\mathbf{\Phi}\in\mathbb{R}^{40 \times 64}$ using the proposed approach 1 with $\alpha=1$ and
$\alpha=0.1$, and using the proposed approach 2. We compare the performance of the proposed
designs with a random Gaussian matrix design and with three iterative designs, namely, Elad's
design~\cite{4359525}, Xu's design~\cite{xu2010optimized} and Sapiro's design~\cite{5061489}.

\begin{figure}[!thb]%
\centering%
\includegraphics[width=0.5\textwidth]{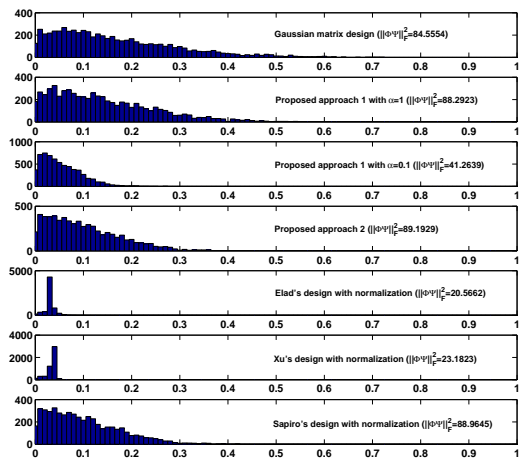}%
\DeclareGraphicsExtensions. \caption{Histogram of the absolute value of the off-diagonal entries of the coherence
matrix.} \label{fig:4-dist_corr}
\end{figure}%

\par
It has been observed that coherence matrices with small off-diagonal entries result in good reconstruction performance
in accordance with the mutual coherence reconstruction condition~\cite{4359525,5061489,xu2010optimized}.
Fig.~\ref{fig:4-dist_corr} shows that the distributions of the off-diagonal entries in both designs based on approach 1
are better than that for the Gaussian matrix design. In particular, note that the design with $\alpha=0.1$ has
off-diagonal entries with smaller absolute value than does the design with $\alpha=1$. However,
$\|\mathbf{\Phi\Psi}\|_F^2=41.2639$ for the $\alpha=0.1$ design is lower than $\|\mathbf{\Phi\Psi}\|_F^2=89.1929$ for
the $\alpha=1$ design - owing to the lower penalty used in the optimization problem in
(\ref{eq:CS_4_optimization_overcomplete}) - and also lower than $\|\mathbf{\Phi\Psi}\|_F^2=84.5554$ for the Gaussian
design. This observation - via the analysis in Section~\uppercase\expandafter{\romannumeral3} - ought to lead to poorer
MSE performance of the design with $\alpha=0.1$ in relation to the design with $\alpha=1$ and also in relation to the
Gaussian design. The distribution of the off-diagonal entries in the design based on approach 2 is also better than the
Gaussian matrix. In addition, the sensing energy of the equivalent sensing matrix $\|\mathbf{\Phi\Psi}\|_F^2=89.1929$
is not reduced compared to the Gaussian matrix design. Elad's and Xu's designs, exhibit good mutual coherence but poor
sensing energy. The attributes of Sapiro's design are equivalent to those of the design based on approach 2. Yet, our
design is non-iterative whereas Sapiro's design follows an iterative procedure.

\par
The reconstruction performance of the proposed designs is further investigated in the following subsections, both in
terms of the MSE of the ideal oracle estimator as well as the MSE of practical estimators.

\subsection{The MSE performance using the oracle estimator}
In this investigation, we evaluate the MSE performance of various designs using the ideal oracle estimator, which has
played a key role in the definition of our designs. The MSE is evaluated by averaging over 1000 trials, where in each
trial we generate randomly a sparse vector with $s$ randomly placed $\pm1$ spikes\footnote{We have also performed this
experiment and the following experiment with sparse vectors where the randomly placed non-zero elements follow a
zero-mean unit-variance Gaussian distribution. Such experiments, which are not reported in view of space limitations,
also demonstrate that our designs outperform other designs in the literature.}. The random dictionary
$\mathbf{\Psi}\in\mathbb{R}^{64 \times 80}$ is generated randomly by drawing its elements from i.i.d. zero mean and
unit variance Gaussian distributions and then normalized to $\|\mathbf{\Psi}\|_F^2=80$. The parameter $\alpha$ is set
to be equal to 1 for the design based on approach 1.

\begin{figure}[!thb]%
\centering%
\includegraphics[width=0.45\textwidth]{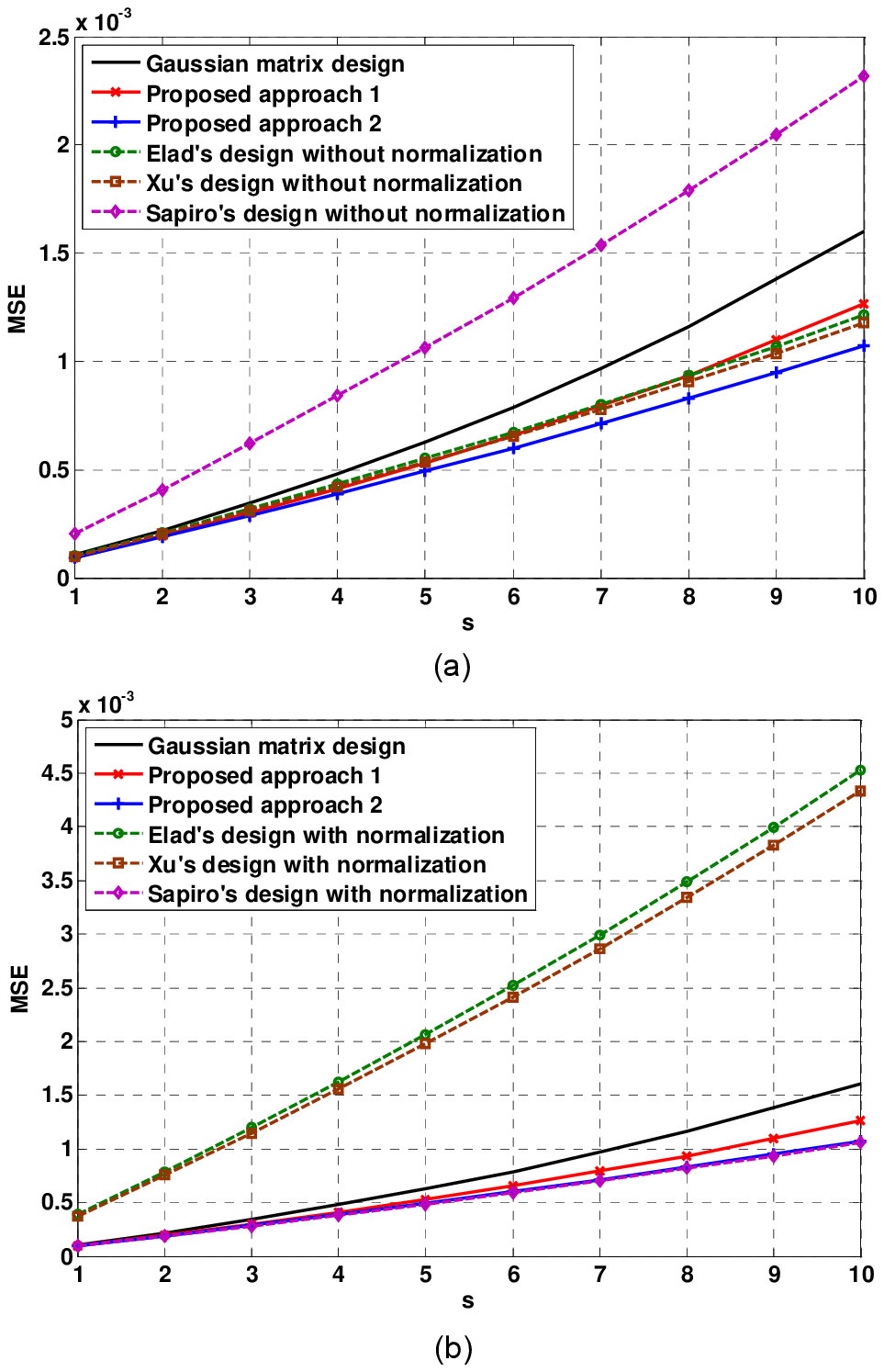}%
\DeclareGraphicsExtensions. \caption{MSE performance of different sensing matrices for the oracle estimator ($m=40$,
$n=64$ $\hat{n}=80$ and $\sigma^2=10^{-4}$). (a) Elad's, Xu's and Sapiro's designs are not normalized; (b) Elad's, Xu's
and Sapiro's designs are normalized.} \label{fig:4-oracle_mse_random}
\end{figure}%
\par
Fig.~\ref{fig:4-oracle_mse_random} illustrates that the performance of our designs compare very well with that of the
best iterative designs. A particularly relevant aspect relates to the sensing matrix normalization of iterative
designs. Sapiro's design works very well with normalization but Elad's and Xu's designs do not. In fact, the MSE
performance of Elad's and Xu's design is worse than that of the random Gaussian design, due to the lower sensing energy
(see Fig.~\ref{fig:4-dist_corr}). The proposed approach 2 has a better MSE performance than approach 1, as the
parameter $\alpha$ of approach 1, which is set empirically, affects the performance.

\subsection{The MSE performance using practical estimators}
In this investigation, we evaluate the MSE performance of various sensing matrix designs using practical estimators,
which include the BPDN, the Dantzig selector and the OMP. As in the previous investigation, the MSE is evaluated by
averaging over 1000 trials, where in each trial we generate randomly a sparse vector with $s$ randomly placed $\pm1$
spikes. The random dictionary $\mathbf{\Psi}\in\mathbb{R}^{64 \times 80}$ is also generated randomly by drawing its
elements from i.i.d. zero mean and unit variance Gaussian distributions and then normalized to
$\|\mathbf{\Psi}\|_F^2=80$. The parameter $\alpha$ is also set to be equal to 1 for the design based on approach 1.

\begin{figure*}[!thb]%
\centering%
\includegraphics[width=1.0\textwidth]{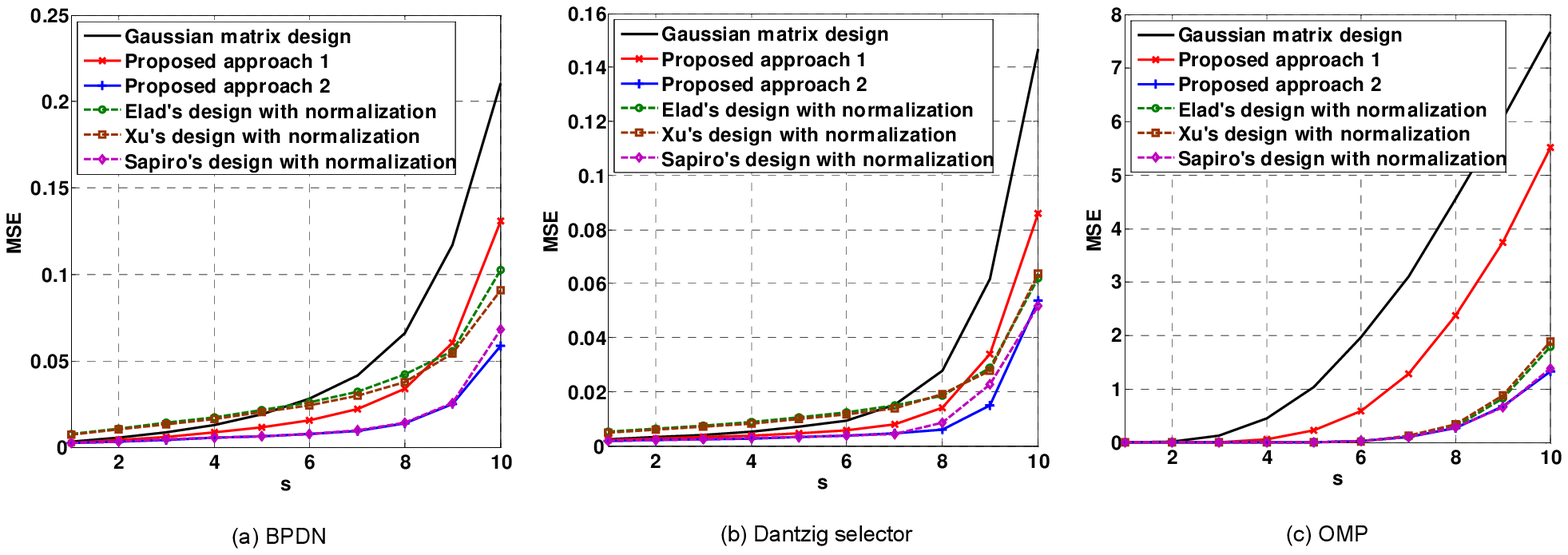}%
\DeclareGraphicsExtensions. \caption{MSE performance of different sensing matrices for (a) the BPDN, (b) the Dantzig
selector, and (c) the OMP ($m=40$, $n=64$ $\hat{n}=80$ and $\sigma^2=10^{-4}$).} \label{fig:4-practical_mse_s}
\end{figure*}%

\par
We first evaluate the MSE performance of various sensing matrix designs for various sparsity levels and for a fixed
number of measurements, $m=40$. Fig.~\ref{fig:4-practical_mse_s} shows that the proposed design approach 1 outperforms
the Gaussian matrix design for all the three estimators. In turn, the proposed design approach 2 outperforms all the
other designs. In fact, this design is very attractive, due to the low computation cost associated with the generation
of the sensing matrix.

\begin{figure*}[!htb]%
\centering%
\includegraphics[width=1.0\textwidth]{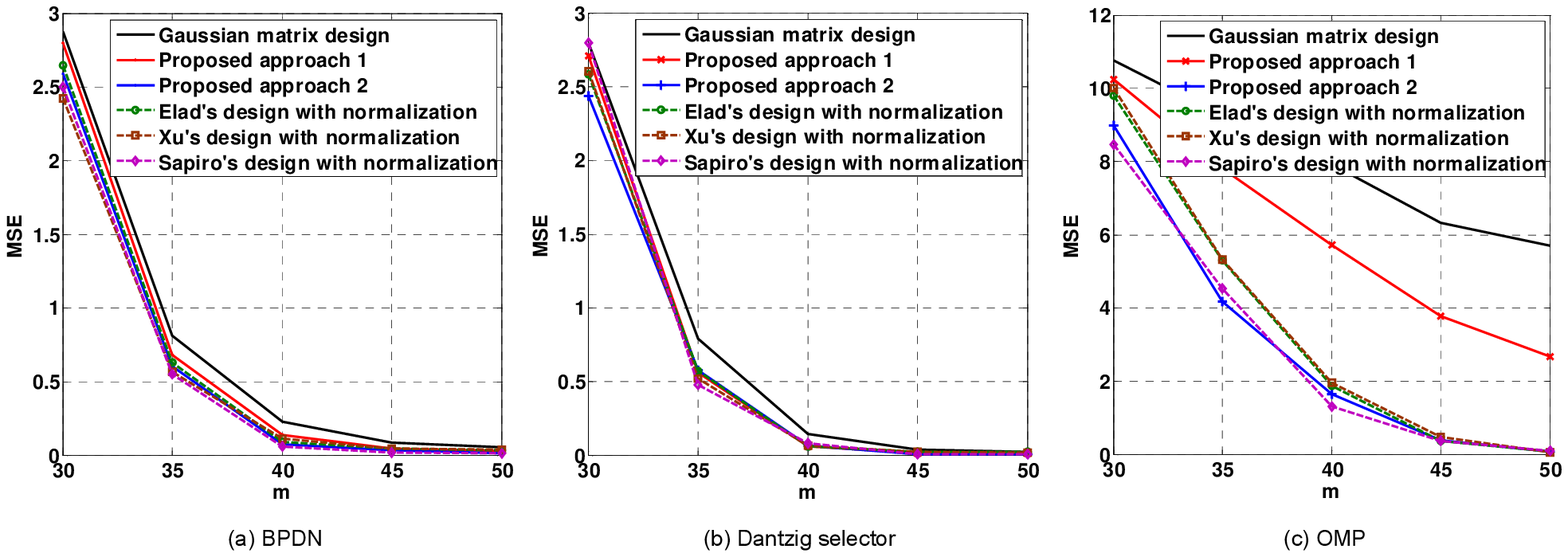}%
\DeclareGraphicsExtensions. \caption{MSE performance of different sensing matrices for (a) the BPDN, (b) the Dantzig
selector, and (c) the OMP ($s=10$, $n=64$ $\hat{n}=80$ and $\sigma^2=10^{-4}$).} \label{fig:4-practical_mse_m}
\end{figure*}%

\par
We now evaluate the MSE performance of various sensing matrix designs for various numbers of measurements and for a
fixed sparsity level $s=10$. Fig.~\ref{fig:4-practical_mse_m} shows once again that the proposed designs outperform the
Gaussian matrix design. We note that the proposed designs improve the reconstruction performance for all the three
estimators, compared to the Gaussian matrix design. The iterative Elad's design, Sapiro's design, and Xu's design,
slightly outperform the proposed designs in some cases, but the computation complexity associated with the generation
of these designs is much higher than that associated with the generation of our design.

\subsection{The reconstruction performance for learned dictionaries in CS imaging}
We now assess the performance of the proposed designs by considering other practical issues. In particular, we consider
real rather than synthetic signals whose representations are typically nearly sparse instead of sparse in some
dictionary. We also consider learned dictionaries rather than random ones\footnote{We note that the dictionary learning
process yields sparse representations that do not necessarily fit the statistical signal model that has been used as a
basis of the sensing matrix design procedure. However, the value of the sensing matrix designs is also justified by the
fact that it also yields observable gains in this scenario.}.

\par
In the experiment, we use the cameraman image of size $256\times 256$ pixels, which is partitioned into $1024$
nonoverlapping patches of size $8\times8$ pixels, i.e., $n=64$. We train a dictionary of size $64\times 81$ for
sparsely representing these nonoverlapping patches by using the K-SVD method~\cite{4011956}. The number of measurements
for each patch is set to be equal to 40 and the measurements are corrupted by additive zero-mean Gaussian noise with
variance $\sigma^2 = 10^{-3}$. We set $\alpha = 1$ for the proposed approach 1. We also use the OMP to reconstruct the
image from its noisy measurements owing to its fast execution. We evaluate performance using the reconstructed signal
to noise ratio (RSNR):
\begin{equation}\label{eq:CS_4_reconstruction_SNR}
\text{RSNR}=\frac{\|\tilde{\mathbf{f}}\|_2}{\|\mathbf{f}-\tilde{\mathbf{f}}\|_2},
\end{equation}
where $\mathbf{f}$ represents the original image and $\tilde{\mathbf{f}}$ represents the reconstructed image.

\begin{figure*}[!thb]%
\centering%
\includegraphics[width=0.8\textwidth]{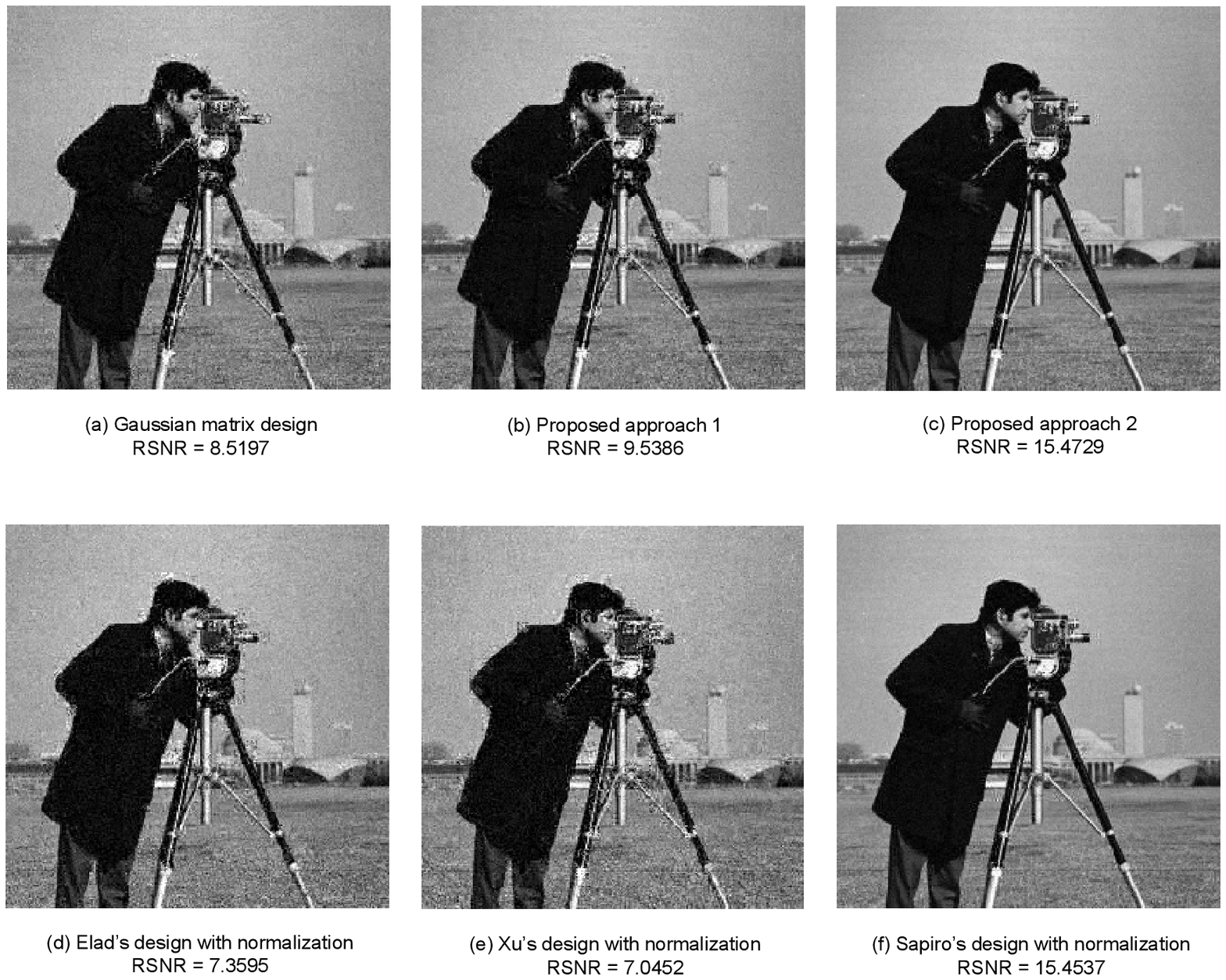}%
\DeclareGraphicsExtensions. \caption{Reconstructed images using a learned basis.} \label{fig:4-pic_perf}
\end{figure*}%
\par
Fig.~\ref{fig:4-pic_perf} demonstrates the higher reconstruction quality and RSNR of our sensing matrix designs in
relation to the random Gaussian matrix design. The proposed approach 2 exhibits the best performance. Sapiro's
iterative design also exhibits a very good performance but Elad's and Xu's iterative designs with normalized sensing
energy exhibit very poor performance, which in fact is worse than that for Gaussian matrix design. Interestingly we
recall that the performance of the proposed two designs compare well to that of Gaussian matrix design for random basis
and exactly sparse signals as shown in Fig.~\ref{fig:4-practical_mse_s} and~\ref{fig:4-practical_mse_m}.

\section{Discussion: Random vs. Optimized Projections}
Recent results~\cite{Candes2012,Ye1953043,6034739} have established that - at least asymptotically
with the signal ambient dimension - no sensing or reconstruction strategy leads to essentially
better performance than random sensing and standard $\ell_1$ based reconstruction. In contrast,
our results indicate that a tight-frame based sensing matrix design can clearly outperform a
random sensing matrix design for low signal ambient dimensions.

\par
It is thus interesting to ask whether our optimized designs can also outperform the random ones with an increase of the
signal ambient dimension. This question is also justified by the fact that the recent contributions in the literature
concentrate on signals that are sparse in the canonical basis rather than signals that are sparse in an overcomplete
dictionary. Interestingly, the numerical analysis reveals that the trends applicable to overcomplete dictionaries can
be distinct from those applicable to the canonical dictionary (and also orthonormal ones).

\par
The experiments also consider randomly generated sparse vectors with $s$ randomly placed $\pm 1$ spikes. We consider
both a random Gaussian sensing matrix design and an optimized sensing matrix design based on approach 2 due to its low
computational cost. The sensing matrix designs are normalized such that $\|\mathbf{\Phi}\|_F^2=n$. We also consider
three distinct dictionaries: i) the canonical basis; ii) a random overcomplete dictionary; and iii) a specified
overcomplete dictionary. The random overcomplete dictionary is generated by drawing its elements randomly in accordance
with i.i.d. zero-mean unit-variance Gaussian distributions. The specified overcomplete dictionary is generated via its
singular value decomposition by taking two randomly generated orthonormal matrices and by taking its positive singular
values $\lambda_1^{\Psi}\geq\ldots\geq\lambda_n^{\Psi}$ such that $\lambda_1^{\Psi}=1$ and
$\frac{\lambda_{i+1}^{\Psi}}{\lambda_{i}^{\Psi}}=0.995$ $(i=1,\ldots,n-1)$. Both overcomplete dictionaries are also
normalized such that $\|\mathbf{\Psi}\|_F^2=\hat{n}$.

\begin{figure}[!t]%
\centering%
\includegraphics[width=0.45\textwidth]{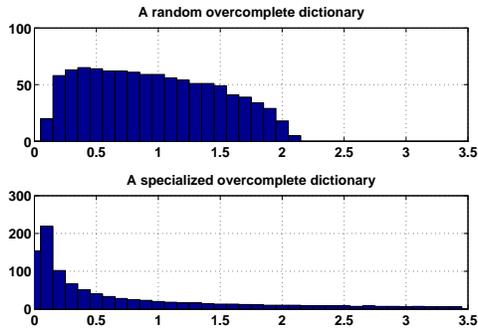}%
\DeclareGraphicsExtensions. \caption{Histogram of the singular values of a random overcomplete dictionary ($n=1000$,
$\hat{n}=1200$ and $\|\mathbf{\Psi}\|_F^2=\hat{n}$) and a specialized overcomplete dictionary ($n=1000$,
$\hat{n}=1200$, $\|\mathbf{\Psi}\|_F^2=\hat{n}$ and $\frac{\lambda_{i+1}^{\Psi}}{\lambda_{i}^{\Psi}}=0.995$).}
\label{fig:sigular_value_dist}
\end{figure}%
\par
The rationale for considering two different overcomplete dictionaries is because it is not entirely clear how to change
the dictionary as the signal ambient dimension is varied\footnote{Note that this issue is not relevant when the signal
dimension is fixed as in the previous experiments (or for the canonical basis).}. Therefore, two overcomplete
dictionaries that exhibit a very different singular value profile as shown in Fig.~\ref{fig:sigular_value_dist}, are
chosen that will allow us to articulate different trends in the experiments.

\par
The MSE performance associated with the various sensing matrix designs is also averaged over 1000 trials. We unveil the
performance trends by showing how the ratio of the average MSE associated with the optimized sensing matrix design to
the average MSE associated with a random sensing matrix design behaves as a function of the signal ambient dimension
for various combinations of ($m$, $s$), both for the Dantzig selector and the oracle estimator. The signal dimension is
restricted to $n=1000$ due to the long execution time of the simulations.

\begin{figure}[!t]%
\centering%
\includegraphics[width=0.45\textwidth]{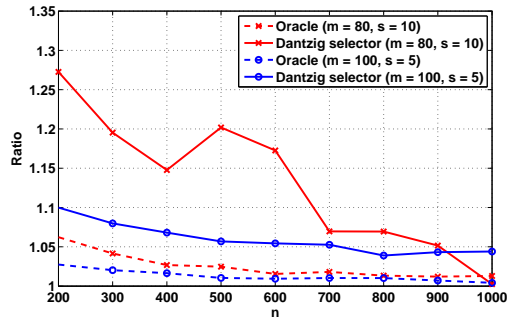}%
\DeclareGraphicsExtensions. \caption{Ratio of the average MSE associated with an optimized sensing
matrix design to that associated with a random Gaussian sensing matrix design for signals that are
sparse in the canonical basis ($\sigma^2=10^{-4}$).} \label{fig:discussion_canonical}
\end{figure}%

\subsection{Case I: Signals that are sparse on the canonical basis}
Fig.~\ref{fig:discussion_canonical} examines how the ratio of the average MSE associated with the
optimized sensing matrix design to the average MSE associated with a random Gaussian sensing
matrix design - which is a tight frame - behaves as a function of the signal dimension. One
observes that the average MSE ratio tends to one with the increase of the signal dimension both
for the oracle estimator and the Dantzig selector. This is due to the fact that a random Gaussian
matrix tends to a tight frame with the increase of $n$ for a fixed $m$~\cite{Rudelson2009}.

\par
It turns out that this result is consistent with the result in~\cite{Candes2012}, where random
sensing matrix designs are demonstrated to be near-optimal (asymptotically) for signals that are
sparse in the canonical basis.

\begin{figure}[!t]%
\centering%
\includegraphics[width=0.45\textwidth]{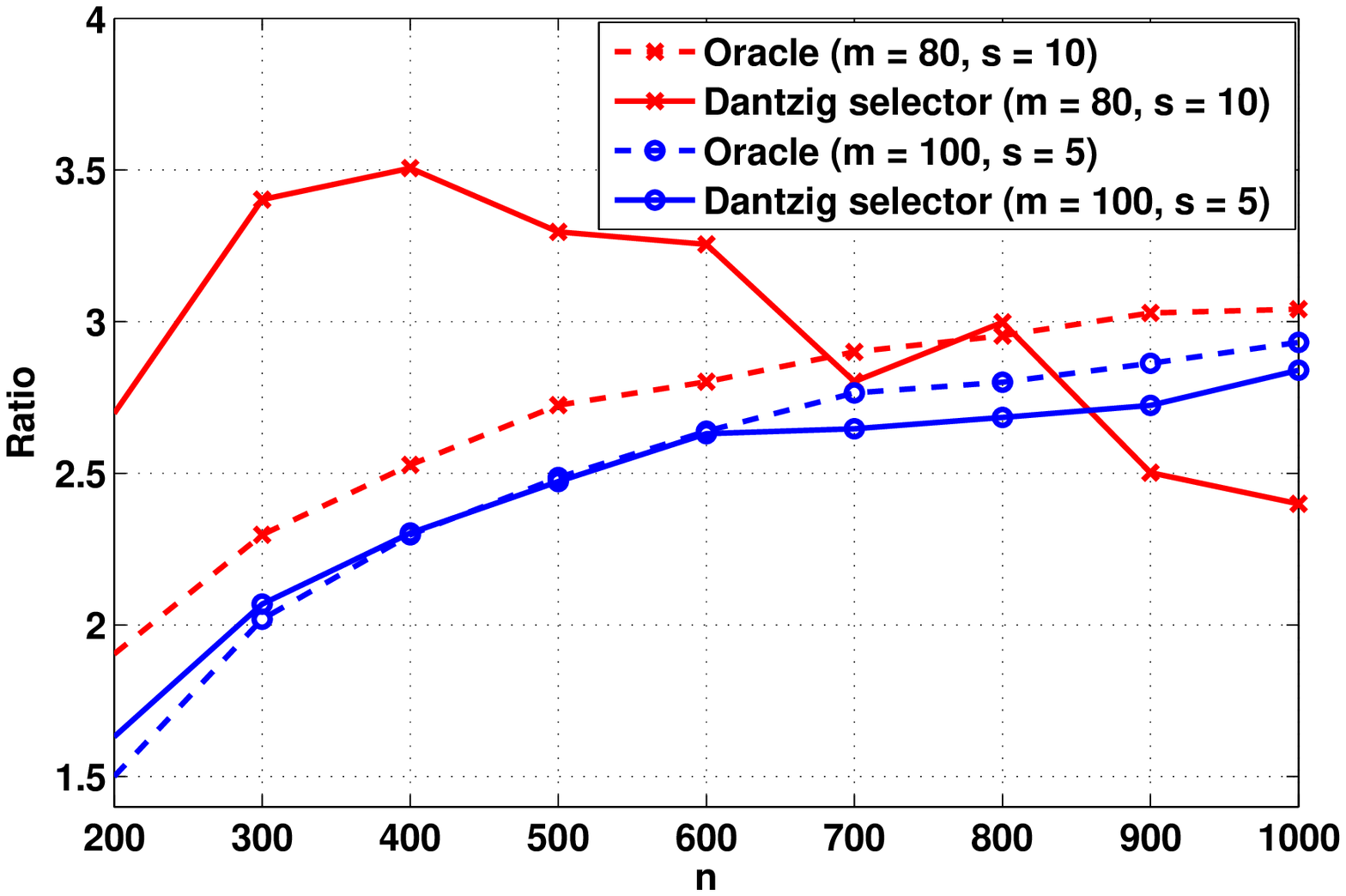}%
\DeclareGraphicsExtensions. \caption{Ratio of the average MSE associated with an optimized sensing
matrix design to that associated with a random Gaussian sensing matrix design for signals that are
sparse in a randomly generated overcomplete dictionarie ($\hat{n}=1.2n$ and $\sigma^2=10^{-4}$).}
\label{fig:discussion_randomDictionary}
\end{figure}%

\begin{figure}[!t]%
\centering%
\includegraphics[width=0.45\textwidth]{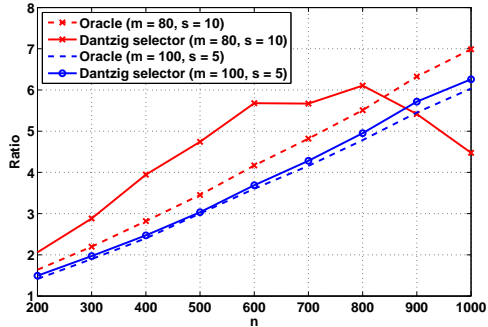}%
\DeclareGraphicsExtensions. \caption{Ratio of the average MSE associated with an optimized sensing
matrix design to that associated with a random Gaussian sensing matrix design for signals that are
sparse in a specified overcomplete dictionary ($\hat{n}=1.2n$ and $\sigma^2=10^{-4}$).}
\label{fig:discussion_specializedDictionary}
\end{figure}%

\subsection{Case II: Signals that are sparse on an overcomplete dictionary}
Figs.~\ref{fig:discussion_randomDictionary} and~\ref{fig:discussion_specializedDictionary} examine how the average MSE
ratio behaves as a function of the signal dimension for the random and specified overcomplete dictionaries,
respectively. One now observes that - and in sharp contrast to the canonical basis scenario - the average MSE ratio
tends to increase with the increase of the signal dimension. This trend is exhibited by the oracle estimator for the
pairs ($m=100$, $s=5$) and ($m=80$, $s=10$). The trend is also exhibited by the Dantzig selector for ($m=100$, $s=5$)
but not for ($m=80$, $s=10$): this exception seems to be due to severe reconstruction errors in view of the fact that
one may not be satisfying the requirement $m=\mathcal{O}(s\log(n/s))$~\cite{1580791,1614066}.

\par
It is relevant though to point out a major difference in the behavior of the trends for the random and specified
overcomplete dictionaries. For the random dictionary, the average MSE ratio appears to saturate with the increase of
the signal dimension: this fact can be justified by noting that not only does the optimized design tends to a tight
frame with the increase of $n$ for a fixed $m$ - because the $m$ largest singular values of a random dictionary tend to
be similar with the increase of $n$ for a fixed $m$ (see also Fig.~\ref{fig:sigular_value_dist}) - but also the random
Gaussian matrix design also tends to a tight frame with the increase of $n$ for a fixed $m$ as discussed previously. In
contrast, for the specified dictionary the average MSE ratio does not appear to saturate with the increase of the
signal dimension.

\begin{figure}[!t]%
\centering%
\includegraphics[width=0.45\textwidth]{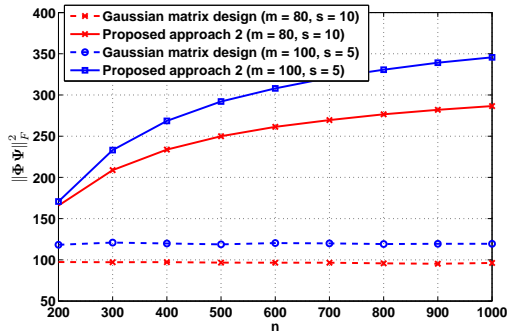}%
\DeclareGraphicsExtensions. \caption{Average sensed energy $\|\mathbf{\Phi \Psi}\|_F^2$ for a random overcomplete
dictionary ($\hat{n}=1.2n$).} \label{fig:discussion_randomDictionary_energy}
\end{figure}%

\begin{figure}[!t]%
\centering%
\includegraphics[width=0.45\textwidth]{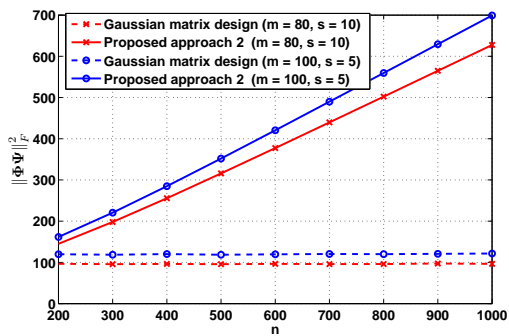}%
\DeclareGraphicsExtensions. \caption{Average sensed energy $\|\mathbf{\Phi \Psi}\|_F^2$ for a specialized overcomplete
dictionary ($\hat{n}=1.2n$, $\|\mathbf{\Psi}\|_F^2=\hat{n}$ and
$\frac{\lambda_{i+1}^{\Psi}}{\lambda_{i}^{\Psi}}=0.995$).} \label{fig:discussion_specializedDictionary_energy}
\end{figure}%

\par
It turns out that such trends can also be partly reconciled with the arguments of the previous sections. In particular,
Figs.~\ref{fig:discussion_randomDictionary_energy} and~\ref{fig:discussion_specializedDictionary_energy} depict how the
average sensed energy (i.e., the energy present at the input to the estimator) behaves as a function of the signal
dimension for the random and the specified overcomplete dictionaries, respectively. Note that the average sensed energy
$\mathbb{E}_{\mathbf{x}}\left(\text{Tr}\left(\mathbf{\Phi \Psi
x}\mathbf{x}^T\mathbf{\Psi}^T\mathbf{\Phi}^T\right)\right)$ corresponds to the equivalent sensing matrix energy
$\|\mathbf{\Phi \Psi}\|_F^2$ in view of the fact that
$\mathbb{E}_{\mathbf{x}}(\mathbf{x}\mathbf{x}^T)=\mathbf{I}_{\hat{n}}$. We would like to emphasize that for both
Figs.~\ref{fig:discussion_randomDictionary_energy} and~\ref{fig:discussion_specializedDictionary_energy} the sensing
matrix designs have been normalized such that $\|\mathbf{\Phi}\|_F^2=n$.

\par
One observes clearly that the optimized designs have the capability to ``sense" higher energy than the random ones in
the presence of overcomplete dictionaries (both the random and the specified overcomplete dictionary) and - via the
analysis in Section~\uppercase\expandafter{\romannumeral3} - potentially have the capability to offer a lower MSE (as
confirmed in Figs.~\ref{fig:discussion_randomDictionary} and~\ref{fig:discussion_specializedDictionary}).
Figs.~\ref{fig:discussion_randomDictionary_energy} and~\ref{fig:discussion_specializedDictionary_energy} also confirm
that for the random dictionary the sensed energy tends to saturate with the increase of the signal dimension but for
the specified dictionary it does not.

\par
We recognize that this analysis is mainly heuristic: a proper understanding of the advantages of designed projections
over random ones in the presence of signals that admit sparse representations in overcomplete dictionaries is beyond
the scope of this article. However, the practical relevance of the overall results - independently of whether or not it
can be crisply shown that optimized projections clearly outperform random ones for high ambient dimensions - is also
associated with the fact that in some applications it is typical to deal with small dimensions. For example, in certain
imaging applications it is standard practice to divide an image into various (possibly overlapping) patches of
typically small dimensions~\cite{4011956}. The results then show that there is indeed significant value in using
optimized projections in lieu of random ones.

\section{Conclusions}
In this paper, we have considered the design of sensing matrices for CS applications. By showing that one ought to set
the equivalent sensing matrix to be equal to a tight frame in order to derive a good MSE performance subject to sensing
energy constraints, we have proposed two sensing matrix designs that are instilled with operational significance. Our
designs also exhibit various advantages in relation to other designs in the literature. In particular, the proposed
designs exhibit MSE performance gains in relation to the conventional random sensing matrix designs as well as other
optimized designs. The proposed designs are also closed-form, and as a result easy to generate, whereas other optimized
designs in the literature are typically iterative.

\section*{Acknowledgment}
The authors would like to thank the anonymous reviewers for their valuable comments and
suggestions that significantly contributed to improving the quality of the paper.

\appendices
\section{Proof of the Proposition~\ref{thm:CS_4_optimization}}
This proof follows the ideas of the proof of Proposition 1 in~\cite{6061944}. Let $\hat{s}\leq s$
be a positive integer. Let also $\mathcal{J}_{\hat{s}}^t\subset\{1,2,\ldots, \hat{n}\}$
($t=1,\ldots,\emph{T}_{\hat{s}}$) denote a support set with cardinality $\hat{s}$, where
$\emph{T}_{\hat{s}}=\binom{\hat{n}}{\hat{s}}=\frac{\hat{n}!}{\hat{s}!(\hat{n}-\hat{s})!}$. We let
$\mathbf{D}_{\mathcal{J}_{\hat{s}}^t}=\mathbf{E}_{\mathcal{J}_{\hat{s}}^t}^T\mathbf{Q}\mathbf{E}_{\mathcal{J}_{\hat{s}}^t}$.
We also let
$\lambda_{\hat{s}}^{\mathcal{J}_{\hat{s}}^t}\geq\ldots\geq\lambda_1^{\mathcal{J}_{\hat{s}}^t}$ be
the eigenvalues of $\mathbf{D}_{\mathcal{J}_{\hat{s}}^t}$. Let
$\text{Pr}(\left\|\mathcal{J}\right\|_0=\hat{s})$ denote the probability that the support size of
$\mathcal{J}$ is $\hat{s}$.

\par
We now note that
\begin{equation}\label{eq:CS_3_optMSE_Proposition_proof_1}
\begin{split}
\sum_{t=1}^{\emph{T}_{\hat{s}}}\text{Tr}\left(\mathbf{D}_{\mathcal{J}_{\hat{s}}^t}\right)
&=\text{Tr}\left(\left(\sum_{t=1}^{\emph{T}_{\hat{s}}}\mathbf{E}_{\mathcal{J}_{\hat{s}}^t}
\mathbf{E}_{\mathcal{J}_{\hat{s}}^t}^T\right)\mathbf{Q}\right)\\
&=\text{Tr}\left(\frac{\hat{s}\emph{T}_{\hat{s}}}{\hat{n}}\mathbf{I}_{\hat{n}}\mathbf{Q}\right)\\
&=\frac{m\hat{s}\emph{T}_{\hat{s}}}{\hat{n}}.
\end{split}
\end{equation}
By the arithmetic mean - harmonic mean inequality, it follows that:
\begin{equation}\label{eq:CS_3_optMSE_Proposition_proof_2}
\sum_{t=1}^{\emph{T}_{\hat{s}}}\sum_{\hat{n}=1}^{\hat{s}}\frac{1}{\lambda_{\hat{n}}^{\mathcal{J}_{\hat{s}}^t}}\geq
\frac{(\hat{s}\emph{T}_{\hat{s}})^2}{\sum_{t=1}^{\emph{T}_{\hat{s}}}\sum_{\hat{n}=1}^{\hat{s}}\lambda_{\hat{n}}^{\mathcal{J}_{\hat{s}}^t}}
=\frac{(\hat{s}\emph{T}_{\hat{s}})^2}{\sum_{t=1}^{\emph{T}_{\hat{s}}}\text{Tr}\left(\mathbf{D}_{\mathcal{J}_{\hat{s}}^t}\right)}
=\frac{\hat{n}\hat{s}\emph{T}_{\hat{s}}}{m},
\end{equation}
where one achieves the lower bound with $\lambda_{\hat{n}}^{\mathcal{J}_{\hat{s}}^t}=\frac{m}{\hat{n}}$
($\hat{n}=1,\ldots,\hat{s}$; $t=1,\ldots,\emph{T}_{\hat{s}}$). This implies immediately that the matrix
$\mathbf{Q}=\frac{m}{\hat{n}}\mathbf{I}_{\hat{n}}$, which is consistent with the constraints, minimizes:
\begin{equation}\label{eq:CS_3_optMSE_Proposition_proof_3}
\mathbb{E}_{\mathcal{J}_{\hat{s}}}\left(\text{Tr}\left(\left(\mathbf{D}_{\mathcal{J}_{\hat{s}}}\right)^{-1}\right)\right)
=\frac{1}{\emph{T}_{\hat{s}}}\sum_{t=1}^{\emph{T}_{\hat{s}}}\sum_{\hat{n}=1}^{\hat{s}}\frac{1}{\lambda_{\hat{n}}^{\mathcal{J}_{\hat{s}}^t}},
\end{equation}
and hence also minimizes:
\begin{equation}\label{eq:CS_3_optMSE_Proposition_proof_4}
\mathbb{E}_{\mathcal{J}}\left(\text{Tr}\left(\left(\mathbf{D}_{\mathcal{J}}\right)^{-1}\right)\right)
=\sum_{\hat{s}=1}^{s}\text{Pr}(\left\|\mathcal{J}\right\|_0=\hat{s})\mathbb{E}_{\mathcal{J}_{\hat{s}}}\left(\text{Tr}\left(\left(\mathbf{D}_{\mathcal{J}_{\hat{s}}}\right)^{-1}\right)\right).
\end{equation}

\section{Proof of the Proposition~\ref{thm:CS_4_solution}}
This proof follows the ideas of the proof of Proposition 2 in~\cite{6061944}. By using the SVD
$\mathbf{A}=\mathbf{U}_\mathbf{A}\mathbf{\Lambda}_\mathbf{A}\mathbf{V}_\mathbf{A}^T$, where
$\mathbf{U}_\mathbf{A}\in\mathbb{R}^{m \times m}$ and $\mathbf{V}_\mathbf{A}\in\mathbb{R}^{\hat{n}
\times \hat{n}}$ are orthonormal matrices, and $\mathbf{\Lambda}_\mathbf{A}\in\mathbb{R}^{m \times
\hat{n}}$ is a matrix whose main diagonal entries ($\lambda^{A}_1\geq \lambda^{A}_2\geq \ldots
\lambda^{A}_m\geq 0$) are the singular values of $\mathbf{A}$, and the off-diagonal entries are
zeros, we pose the convex optimization problem:
\begin{equation}\label{eq:CS_3_optmatrix_Proposition_proof_2}
\begin{split}
\min_{\lambda^{A}_1,\ldots,\lambda^{A}_m}&\ \sum_{i=1}^{m}\left(\left(\lambda^{A}_i\right)^2-\frac{m}{\hat{n}}\right)^2\\
\text{s.t.}& \ \sum_{i=1}^{m}\left(\lambda^{A}_i\right)^2=m,\ \lambda^{A}_i\geq0\ (i=1,2,\ldots,m).
\end{split}
\end{equation}
which, in view of the fact that
$\left\|\mathbf{A}^T\mathbf{A}-\frac{m}{\hat{n}}\mathbf{I}_{\hat{n}}\right\|_F^2=\sum_{i=1}^{m}\left(\left(\lambda^{A}_i\right)^2-\frac{m}{\hat{n}}\right)^2$
and $\texttt{Tr}\left(\mathbf{A}^T\mathbf{A}\right)=\sum_{i=1}^{m}\left(\lambda^{A}_i\right)^2$, leads to the solution
of (\ref{eq:CS_4_optmatrix_Proposition}). Since the solution of (\ref{eq:CS_3_optmatrix_Proposition_proof_2}) is
$\lambda^{A}_i=1$ ($i=1,2,\ldots,m$), it follows that any Parseval tight frame is the solution of
(\ref{eq:CS_4_optmatrix_Proposition}).

\section{Tight frames and StRIP}
Another benefit of tight frames - more precisely, unit-norm tight frames - is its relation to the weaker version of the
RIP, namely the StRIP. The StRIP, which has been proposed by Calderbank et al.~\cite{5419073}, can be used to evaluate
the expected-case performance of CS, whereas the RIP is a worst-case performance indicator, as is the mutual coherence.
The StRIP guarantees successful reconstruction of all but an exponentially small fraction of $s$ sparse signals. The
definition of StRIP uses a probability criterion to replace the hard requirement demanded in the definition of the RIP.

\begin{definition}
A matrix $\mathbf{A}\in\mathbb{R}^{m\times \hat{n}}$ is said to be an $(s,\delta,\eta)$-StRIP matrix if for $s$ sparse
vectors $\mathbf{x}\in\mathbb{R}^{\hat{n}}$ the inequalities
\begin{equation}\label{eq:CS_3_StRIP}
(1-\delta)\|\mathbf{x}\|_{2}^2\leq\left\|\mathbf{Ax}\right\|_{2}^2\leq (1+\delta)\|\mathbf{x}\|_{2}^2
\end{equation}
hold with probability exceeding $1-\eta$ with respect to a uniform distribution of the vectors $\mathbf{x}$ among all
$s$ sparse vectors in $\mathbb{R}^{\hat{n}}$ having the same fixed magnitudes.
\end{definition}

\par
Calderbank et al.~\cite{5419073} demonstrate that deterministic sensing matrices are StRIP matrices if they satisfy all
of the following criteria:
\begin{enumerate}
  \item The rows of $\mathbf{A}$ are orthogonal and all the row sums are zero, i.e., $\sum_{i=1}^{\hat{n}} \mathbf{a}_i=\mathbf{0}$;
  \item The columns of $\mathbf{A}$ form a group under point-wise multiplication;
  \item There is one column of $\mathbf{A}$ equal to $\mathbf{1}$, which can be assumed as the first column. For all
  $i\in\{2,\ldots,\hat{n}\}$, $\|\mathbf{a}_i\|_2^2\leq m^{2-\beta}$, where $0<\beta\leq 1$.
\end{enumerate}
A large class of matrices, including discrete chirp sensing matrices, Bose, Chaudhuri, and Hocquenghem (BCH) sensing
matrices, Kerdock, Delsarte-Goethals and second order Reed Muller sensing matrices, satisfy these criteria, and thus
are StRIP matrices. In~\cite{5419073}, they prove that the RIP of these matrices is satisfied with a probability
exceeding $1-\mathcal{O}\left(\exp\left(-\frac{\delta^2m^\beta}{s}\right)\right)$. However, a unit norm tight frame
does not necessarily satisfy these criteria. For example, an orthonormal matrix, which is also a unit norm tight frame
with $m=\hat{n}$, does not necessarily satisfy $\sum_{i=1}^{\hat{n}} \mathbf{a}_i=\mathbf{0}$.

\par
The following Proposition demonstrates that a unit-norm tight frame is also a StRIP matrix

\begin{proposition}\label{thm:CS_4_StRIP_Tropp}
Let $\mathbf{A}\in\mathbb{R}^{m\times \hat{n}}$ be a unit norm tight frame with mutual coherence equal to $\mu$. For
any $s$ sparse vectors $\mathbf{x}\in\mathbb{R}^{\hat{n}}$, the RIP holds with probability:
\begin{equation}\label{eq:CS_3_StRIP_Tropp}
\mathbb{P}\left(\left|\left\|\mathbf{Ax}\right\|_{2}^2-\left\|\mathbf{x}\right\|_2^2\right|\leq
\delta\left\|\mathbf{x}\right\|_2^2\right)>1-(s/2)^{-\frac{\left(0.3894\delta-\frac{s}{m}\right)^2}{36\mu^2s\log_e
(1+s/2)}},
\end{equation}
where $\sqrt{237.42\mu^2s\log_e (1+s/2)}+\frac{2.57s}{m}\leq\delta< 1$.
\end{proposition}

\begin{IEEEproof}
Let $\mathbf{A}_{\mathcal{J}_s}\in\mathbb{R}^{m\times s}$ be an $s$-column submatrix of
$\mathbf{A}\in\mathbb{R}^{m\times \hat{n}}$ ($s<m\leq \hat{n}$), where $\mathcal{J}_s\subset\{1,\ldots,\hat{n}\}$
denotes a support set with cardinality $s$. Let
$\lambda_1^{\mathbf{A}_{\mathcal{J}_s}^T\mathbf{A}_{\mathcal{J}_s}}\geq\ldots\geq\lambda_s^{\mathbf{A}_{\mathcal{J}_s}^T\mathbf{A}_{\mathcal{J}_s}}\geq0$
be the eigenvalues of the positive semi-definite matrix $\mathbf{A}_{\mathcal{J}_s}^T\mathbf{A}_{\mathcal{J}_s}$. We
have that the maximum eigenvalue $\lambda_1^{\mathbf{A}_{\mathcal{J}_s}^T\mathbf{A}_{\mathcal{J}_s}}$ and minimum
eigenvalue $\lambda_s^{\mathbf{A}_{\mathcal{J}_s}^T\mathbf{A}_{\mathcal{J}_s}}$ of
$\mathbf{A}_{\mathcal{J}_s}^T\mathbf{A}_{\mathcal{J}_s}$ are given by
\begin{equation}\label{eq:CS_App_eigenvalue_max}
\lambda_1^{\mathbf{A}_{\mathcal{J}_s}^T\mathbf{A}_{\mathcal{J}_s}}=\max_{\mathbf{z}\neq\mathbf{0}}\frac{\|\mathbf{A}_{\mathcal{J}_s}\mathbf{z}\|_2^2}{\|\mathbf{z}\|_2^2},
\end{equation}
and
\begin{equation}\label{eq:CS_App_eigenvalue_min}
\lambda_s^{\mathbf{A}_{\mathcal{J}_s}^T\mathbf{A}_{\mathcal{J}_s}}=\min_{\mathbf{z}\neq\mathbf{0}}\frac{\|\mathbf{A}_{\mathcal{J}_s}\mathbf{z}\|_2^2}{\|\mathbf{z}\|_2^2},
\end{equation}
where $\mathbf{z}\in\mathbb{R}^s$. Therefore, we have
\begin{equation}\label{eq:CS_App_rip_1}
\lambda_s^{\mathbf{A}_{\mathcal{J}_s}^T\mathbf{A}_{\mathcal{J}_s}}\|\mathbf{z}\|_2^2\leq\|\mathbf{A}_{\mathcal{J}_s}\mathbf{z}\|_2^2\leq\lambda_1^{\mathbf{A}_{\mathcal{J}_s}^T\mathbf{A}_{\mathcal{J}_s}}\|\mathbf{z}\|_2^2,
\end{equation}
or
\begin{equation}\label{eq:CS_App_rip_2}
(\lambda_s^{\mathbf{A}_{\mathcal{J}_s}^T\mathbf{A}_{\mathcal{J}_s}}-1)\|\mathbf{z}\|_2^2\leq\|\mathbf{A}_{\mathcal{J}_s}\mathbf{z}\|_2^2-\|\mathbf{z}\|_2^2\leq(\lambda_1^{\mathbf{A}_{\mathcal{J}_s}^T\mathbf{A}_{\mathcal{J}_s}}-1)\|\mathbf{z}\|_2^2,
\end{equation}
for any $\mathbf{z}\in\mathbb{R}^s$. By defining
$\delta_{\mathcal{J}_s}=\max\left(\left|\lambda_1^{\mathbf{A}_{\mathcal{J}_s}^T\mathbf{A}_{\mathcal{J}_s}}-1\right|,\left|\lambda_s^{\mathbf{A}_{\mathcal{J}_s}^T\mathbf{A}_{\mathcal{J}_s}}-1\right|\right)$,
it follows that
\begin{equation}\label{eq:CS_App_rip_3}
\left|\|\mathbf{A}_{\mathcal{J}_s}\mathbf{z}\|_2^2-\|\mathbf{z}\|_2^2\right|\leq
\delta_{\mathcal{J}_s}\|\mathbf{z}\|_2^2.
\end{equation}
We can immediately derive that the RIC satisfies $\delta_s=\max_{\mathcal{J}_s}\ \delta_{\mathcal{J}_s}$ by comparing
(\ref{eq:CS_App_rip_3}) with (\ref{eq:CS_2_RIP}).

\par
The following theorem, which has been proved by Tropp in~\cite{Tropp20081}, defines a probability bound for
$\delta_{\mathcal{J}_s}$.

\begin{theorem}{Theorem}\label{thm:CS_App_StRIP_Tropp}
Let $\mathbf{A}\in\mathbb{R}^{m\times \hat{n}}$ ($m<\hat{n}$) be a matrix whose columns have unit norm, i.e.,
$\|\mathbf{a}_i\|_2=1$ for all $i\in\{1,\ldots,\hat{n}\}$, $\mathbf{A}_{\mathcal{J}_s}\in\mathbb{R}^{m\times s}$
($s<m$) be a random $s$-column submatrix of $\mathbf{A}$ with a support $\mathcal{J}_s\subset\{1,\ldots,\hat{n}\}$ of
cardinality $s$, and $\mu$ be the mutual coherence of $\mathbf{A}$. Suppose that
\begin{equation}\label{eq:CS_App_StRIP_Tropp_1}
\sqrt{144\mu^2s\log_e (1+s/2)\rho}+\frac{2s}{\hat{n}}\|\mathbf{A}\|^2\leq e^{-0.25}\delta,
\end{equation}
where $\rho\geq1$ and $0<\delta<1$. Then
\begin{equation}\label{eq:CS_App_StRIP_Tropp_2}
\text{Pr}\left(\delta_{\mathcal{J}_s}\geq \delta\right)\leq (s/2)^{-\rho}.
\end{equation}
\end{theorem}

\par
We now use the fact that $\mathbf{A}$ is a unit norm tight frame with frame bound equal to $\frac{\hat{n}}{m}$, so that
$\|\mathbf{A}\|^2=\frac{\hat{n}}{m}$. We then rewrite (\ref{eq:CS_App_StRIP_Tropp_1}) to be
\begin{equation}\label{eq:CS_App_StRIP_Tropp_3}
1\leq\rho\leq\frac{\left(e^{-0.25}\delta-\frac{2s}{m}\right)^2}{144\mu^2s\log_e (1+s/2)},
\end{equation}
where $\sqrt{144e^{0.5}\mu^2s\log_e (1+s/2)}+\frac{2e^{0.25}s}{m}\leq\delta< 1$. Since the inequality
(\ref{eq:CS_App_StRIP_Tropp_2}) holds for any $\rho$ satisfying (\ref{eq:CS_App_StRIP_Tropp_1}), we have that for a
random set $\mathcal{J}_s$, the inequality (\ref{eq:CS_App_rip_3}) leads to the probability bound
\begin{equation}\label{eq:CS_App_StRIP_Tropp_4}
\begin{split}
\mathbb{P}\left(\left|\|\mathbf{A}_{\mathcal{J}_s}\mathbf{z}\|_2^2-\|\mathbf{z}\|_2^2\right|
\leq \delta\|\mathbf{z}\|_2^2\right)\geq &\mathbb{P}\left(\delta_{\mathcal{J}_s}\leq \delta\right)\\
> &1-(s/2)^{-\frac{\left(0.3894\delta-\frac{s}{m}\right)^2}{36\mu^2s\log_e (1+s/2)}},
\end{split}
\end{equation}
when $\sqrt{237.42\mu^2s\log_e (1+s/2)}+\frac{2.57s}{m}\leq\delta< 1$.
\end{IEEEproof}

\par
\emph{Remark 1}: The mutual coherence $\mu$ plays as an important role in this probability bound. The mutual coherence
of various unit norm tight frames could be different, and its distribution is unknown. However, the mutual coherence is
fixed for some specific unit norm tight frames. For example, the Fourier-Dirac tight frame has mutual coherence
$\mu=\frac{1}{\sqrt{m}}$~\cite{Tropp20081}, and the equiangular tight frame has mutual coherence
$\mu=\sqrt{\frac{\hat{n}-m}{m(\hat{n}-1)}}$~\cite{Strohmer2003257}.

\par
\emph{Remark 2}: In~\cite{5419073}, the authors conclude that the RIP is satisfied with a probability exceeding
$1-\mathcal{O}\left(\exp\left(-\frac{\delta^2m^\beta}{s}\right)\right)$ ($0<\beta\leq 1$) for a large class of
matrices, including discrete chirp sensing matrices, Bose, Chaudhuri, and Hocquenghem (BCH) sensing matrices, Kerdock,
Delsarte-Goethals and second order Reed Muller sensing matrices. According to Proposition~\ref{thm:CS_4_StRIP_Tropp},
we can conclude that the RIP holds with a probability that exceeds
$1-\mathcal{O}\left(\exp\left(-\frac{\delta^2m}{s}\right)\right)$ for Fourier-Dirac tight frames and the equiangular
tight frame, so that these tight frames exhibit better quality in terms of the StRIP in relation to the sensing
matrices in~\cite{5419073}.

\par
\emph{Remark 3}: Proposition~\ref{thm:CS_4_StRIP_Tropp} requires unit norm tight frames with frame bound
$\frac{\hat{n}}{m}$. It turns out that, one can scale a unit norm tight frame via
$\mathbf{A}=\frac{\sqrt{m}\mathbf{A}}{\|\mathbf{A}\|_F}$, which leads to a Parseval tight frame with an equal column
norm, in order to achieve the frame bound equal to 1 used in the paper. In fact, scaling the unit norm tight frames
does not change the matrix structure, only the sensing energy.




%
%

\ifCLASSOPTIONcaptionsoff
  \newpage
\fi



%

\bibliographystyle{IEEEtran}
\bibliography{IEEEabrv,bib_paper}

\begin{thebibliography}{10}
\providecommand{\url}[1]{#1}
\csname url@samestyle\endcsname
\providecommand{\newblock}{\relax}
\providecommand{\bibinfo}[2]{#2}
\providecommand{\BIBentrySTDinterwordspacing}{\spaceskip=0pt\relax}
\providecommand{\BIBentryALTinterwordstretchfactor}{4}
\providecommand{\BIBentryALTinterwordspacing}{\spaceskip=\fontdimen2\font plus
\BIBentryALTinterwordstretchfactor\fontdimen3\font minus
  \fontdimen4\font\relax}
\providecommand{\BIBforeignlanguage}[2]{{%
\expandafter\ifx\csname l@#1\endcsname\relax
\typeout{** WARNING: IEEEtran.bst: No hyphenation pattern has been}%
\typeout{** loaded for the language `#1'. Using the pattern for}%
\typeout{** the default language instead.}%
\else
\language=\csname l@#1\endcsname
\fi
#2}}
\providecommand{\BIBdecl}{\relax}
\BIBdecl

\bibitem{1580791}
E.~Cand{\`e}s, J.~Romberg, and T.~Tao, ``Robust uncertainty principles: exact
  signal reconstruction from highly incomplete frequency information,''
  \emph{Information Theory, IEEE Transactions on}, vol.~52, no.~2, pp. 489 --
  509, Feb. 2006.

\bibitem{1614066}
D.~Donoho, ``Compressed sensing,'' \emph{Information Theory, IEEE Transactions
  on}, vol.~52, no.~4, pp. 1289 --1306, Apr. 2006.

\bibitem{lustig2007sparse}
M.~Lustig, D.~Donoho, and J.~Pauly, ``{Sparse MRI: The application of
  compressed sensing for rapid MR imaging},'' \emph{Magnetic Resonance in
  Medicine}, vol.~58, no.~6, pp. 1182--1195, 2007.

\bibitem{6287532}
W.~Chen, M.~Rodrigues, and I.~Wassell, ``A fr{\'e}chet mean approach for
  compressive sensing date acquisition and reconstruction in wireless sensor
  networks,'' \emph{Wireless Communications, IEEE Transactions on}, vol.~11,
  no.~10, pp. 3598 --3606, Oct. 2012.

\bibitem{5437428}
R.~Baraniuk, V.~Cevher, M.~Duarte, and C.~Hegde, ``Model-based compressive
  sensing,'' \emph{Information Theory, IEEE Transactions on}, vol.~56, no.~4,
  pp. 1982 --2001, Apr. 2010.

\bibitem{4524050}
S.~Ji, Y.~Xue, and L.~Carin, ``Bayesian compressive sensing,'' \emph{Signal
  Processing, IEEE Transactions on}, vol.~56, no.~6, pp. 2346 --2356, Jun.
  2008.

\bibitem{hegde2012signal}
C.~Hegde and R.~Baraniuk, ``Signal recovery on incoherent manifolds,''
  \emph{Information Theory, IEEE Transactions on}, vol.~58, no.~12, pp. 7204
  --7214, Dec. 2012.

\bibitem{5559508}
M.~Chen, J.~Silva, J.~Paisley, C.~Wang, D.~Dunson, and L.~Carin, ``Compressive
  sensing on manifolds using a nonparametric mixture of factor analyzers:
  Algorithm and performance bounds,'' \emph{Signal Processing, IEEE
  Transactions on}, vol.~58, no.~12, pp. 6140 --6155, Dec. 2010.

\bibitem{5771110}
X.~Ding, L.~He, and L.~Carin, ``Bayesian robust principal component analysis,''
  \emph{Image Processing, IEEE Transactions on}, vol.~20, no.~12, pp. 3419
  --3430, Dec. 2011.

\bibitem{6094210}
C.~Fu, X.~Ji, and Q.~Dai, ``Adaptive compressed sensing recovery utilizing the
  property of signal's autocorrelations,'' \emph{Image Processing, IEEE
  Transactions on}, vol.~21, no.~5, pp. 2369 --2378, May 2012.

\bibitem{5887383}
Z.~Zhang and B.~Rao, ``Sparse signal recovery with temporally correlated source
  vectors using sparse bayesian learning,'' \emph{Selected Topics in Signal
  Processing, IEEE Journal of}, vol.~5, no.~5, pp. 912 --926, Sept. 2011.

\bibitem{6155613}
T.~Peleg, Y.~Eldar, and M.~Elad, ``Exploiting statistical dependencies in
  sparse representations for signal recovery,'' \emph{Signal Processing, IEEE
  Transactions on}, vol.~60, no.~5, pp. 2286 --2303, May 2012.

\bibitem{danielyan2008image}
A.~Danielyan, A.~Foi, V.~Katkovnik, and K.~Egiazarian, ``Image upsampling via
  spatially adaptive block-matching filtering,'' in \emph{Proc. of 16th
  European Signal Processing Conference, EUSIPCO2008}, Aug. 2008.

\bibitem{4379013}
K.~Egiazarian, A.~Foi, and V.~Katkovnik, ``Compressed sensing image
  reconstruction via recursive spatially adaptive filtering,'' in \emph{Proc.
  IEEE International Conference on Image Processing, ICIP 2007.}, Sept. 2007,
  pp. 549 -- 552.

\bibitem{danielyan2010spatially}
A.~Danielyan, A.~Foi, V.~Katkovnik, and K.~Egiazarian, ``Spatially adaptive
  filtering as regularization in inverse imaging: Compressive sensing,
  super-resolution, and upsampling,'' \emph{Super-Resolution Imaging (P.
  Milanfar, ed.), CRC Press / Taylor $\&$ Francis}, pp. 123--153, Sept. 2010.

\bibitem{4359525}
M.~Elad, ``Optimized projections for compressed sensing,'' \emph{Signal
  Processing, IEEE Transactions on}, vol.~55, no.~12, pp. 5695 --5702, Dec.
  2007.

\bibitem{5061489}
J.~Duarte-Carvajalino and G.~Sapiro, ``Learning to sense sparse signals:
  Simultaneous sensing matrix and sparsifying dictionary optimization,''
  \emph{Image Processing, IEEE Transactions on}, vol.~18, no.~7, pp. 1395
  --1408, Jul. 2009.

\bibitem{xu2010optimized}
J.~Xu, Y.~Pi, and Z.~Cao, ``Optimized projection matrix for compressive
  sensing,'' \emph{EURASIP J. Adv. Signal Process}, vol.~43, pp. 1--8, Feb.
  2010.

\bibitem{5872076}
L.~Zelnik-Manor, K.~Rosenblum, and Y.~Eldar, ``Sensing matrix optimization for
  block-sparse decoding,'' \emph{Signal Processing, IEEE Transactions on},
  vol.~59, no.~9, pp. 4300 --4312, Sept. 2011.

\bibitem{carson2012communications}
W.~Carson, M.~Chen, M.~Rodrigues, R.~Calderbank, and L.~Carin,
  ``Communications-inspired projection design with application to compressive
  sensing,'' \emph{SIAM Journal on Imaging Sciences}, vol.~5, no.~4, pp.
  1185--1212, 2012.

\bibitem{1564423}
D.~Donoho, M.~Elad, and V.~Temlyakov, ``Stable recovery of sparse overcomplete
  representations in the presence of noise,'' \emph{Information Theory, IEEE
  Transactions on}, vol.~52, no.~1, pp. 6 -- 18, Jan. 2006.

\bibitem{5419073}
R.~Calderbank, S.~Howard, and S.~Jafarpour, ``Construction of a large class of
  deterministic sensing matrices that satisfy a statistical isometry
  property,'' \emph{Selected Topics in Signal Processing, IEEE Journal of},
  vol.~4, no.~2, pp. 358 --374, Apr. 2010.

\bibitem{candes2007dantzig}
E.~Cand{\`e}s and T.~Tao, ``{The Dantzig selector: Statistical estimation when
  p is much larger than n},'' \emph{The Annals of Statistics}, vol.~35, no.~6,
  pp. 2313--2351, 2007.

\bibitem{6061944}
W.~Chen, M.~Rodrigues, and I.~Wassell, ``On the use of unit-norm tight frames
  to improve the average mse performance in compressive sensing applications,''
  \emph{Signal Processing Letters, IEEE}, vol.~19, no.~1, pp. 8 --11, Jan.
  2012.

\bibitem{rauhut2008compressed}
H.~Rauhut, K.~Schnass, and P.~Vandergheynst, ``Compressed sensing and redundant
  dictionaries,'' \emph{Information Theory, IEEE Transactions on}, vol.~54,
  no.~5, pp. 2210--2219, May 2008.

\bibitem{Candes201159}
E.~J. Cand{\`e}s, Y.~C. Eldar, D.~Needell, and P.~Randall, ``Compressed sensing
  with coherent and redundant dictionaries,'' \emph{Applied and Computational
  Harmonic Analysis}, vol.~31, no.~1, pp. 59 -- 73, 2011.

\bibitem{davenport2012signal}
M.~Davenport, D.~Needell, and M.~Wakin, ``Signal space cosamp for sparse
  recovery with redundant dictionaries,'' \emph{arXiv preprint
  arXiv:1208.0353}, 2012.

\bibitem{553473}
F.~Champagnat, Y.~Goussard, and J.~Idier, ``Unsupervised deconvolution of
  sparse spike trains using stochastic approximation,'' \emph{Signal
  Processing, IEEE Transactions on}, vol.~44, no.~12, pp. 2988 --2998, Dec.
  1996.

\bibitem{5072251}
H.~Zayyani, M.~Babaie-Zadeh, and C.~Jutten, ``An iterative bayesian algorithm
  for sparse component analysis in presence of noise,'' \emph{Signal
  Processing, IEEE Transactions on}, vol.~57, no.~11, pp. 4378 --4390, Nov.
  2009.

\bibitem{5398963}
N.~Dobigeon and J.-Y. Tourneret, ``Bayesian orthogonal component analysis for
  sparse representation,'' \emph{Signal Processing, IEEE Transactions on},
  vol.~58, no.~5, pp. 2675 --2685, May 2010.

\bibitem{5484983}
R.~Gribonval and K.~Schnass, ``Dictionary identification-sparse
  matrix-factorization via $\ell_1$-minimization,'' \emph{Information Theory,
  IEEE Transactions on}, vol.~56, no.~7, pp. 3523 --3539, Jul. 2010.

\bibitem{1542488}
F.~Labeau, J.-C. Chiang, M.~Kieffer, P.~Duhamel, L.~Vandendorpe, and B.~Macq,
  ``Oversampled filter banks as error correcting codes: theory and impulse
  noise correction,'' \emph{Signal Processing, IEEE Transactions on}, vol.~53,
  no.~12, pp. 4619 -- 4630, Dec. 2005.

\bibitem{hyder2010improved}
M.~Hyder and K.~Mahata, ``An improved smoothed $\ell_0$ approximation algorithm
  for sparse representation,'' \emph{Signal Processing, IEEE Transactions on},
  vol.~58, no.~4, pp. 2194--2205, Apr. 2010.

\bibitem{5930380}
C.~Soussen, J.~Idier, D.~Brie, and J.~Duan, ``From bernoulli-gaussian
  deconvolution to sparse signal restoration,'' \emph{Signal Processing, IEEE
  Transactions on}, vol.~59, no.~10, pp. 4572 --4584, Oct. 2011.

\bibitem{chen2001atomic}
S.~Chen, D.~Donoho, and M.~Saunders, ``{Atomic decomposition by basis
  pursuit},'' \emph{SIAM review}, vol.~43, no.~1, pp. 129--159, 2001.

\bibitem{candes2008restricted}
E.~Cand{\`e}s, ``{The restricted isometry property and its implications for
  compressed sensing},'' \emph{Comptes rendus-Math{\'e}matique}, vol. 346, no.
  9-10, pp. 589--592, 2008.

\bibitem{1542412}
E.~Cand{\`e}s and T.~Tao, ``Decoding by linear programming,'' \emph{Information
  Theory, IEEE Transactions on}, vol.~51, no.~12, pp. 4203 -- 4215, Dec. 2005.

\bibitem{4801665}
S.~Kim and C.~Yoo, ``Underdetermined blind source separation based on subspace
  representation,'' \emph{Signal Processing, IEEE Transactions on}, vol.~57,
  no.~7, pp. 2604 --2614, Jul. 2009.

\bibitem{1300802}
V.~Bostanov, ``Bci competition 2003-data sets ib and iib: feature extraction
  from event-related brain potentials with the continuous wavelet transform and
  the t-value scalogram,'' \emph{Biomedical Engineering, IEEE Transactions on},
  vol.~51, no.~6, pp. 1057 --1061, Jun. 2004.

\bibitem{847906}
G.~Yen and K.-C. Lin, ``Wavelet packet feature extraction for vibration
  monitoring,'' \emph{Industrial Electronics, IEEE Transactions on}, vol.~47,
  no.~3, pp. 650 --667, Jun. 2000.

\bibitem{5428818}
Z.~Ben-Haim and Y.~Eldar, ``The cram{\`e}r-rao bound for estimating a sparse
  parameter vector,'' \emph{Signal Processing, IEEE Transactions on}, vol.~58,
  no.~6, pp. 3384 --3389, Jun. 2010.

\bibitem{5483095}
Z.~Ben-Haim, Y.~Eldar, and M.~Elad, ``Coherence-based performance guarantees
  for estimating a sparse vector under random noise,'' \emph{Signal Processing,
  IEEE Transactions on}, vol.~58, no.~10, pp. 5030 --5043, Oct. 2010.

\bibitem{4063557}
E.~G. Larsson and Y.~Selen, ``Linear regression with a sparse parameter
  vector,'' \emph{Signal Processing, IEEE Transactions on}, vol.~55, no.~2, pp.
  451 -- 460, Feb. 2007.

\bibitem{schniter2009fast}
P.~Schniter, L.~Potter, and J.~Ziniel, ``Fast bayesian matching pursuit: Model
  uncertainty and parameter estimation for sparse linear models,'' \emph{OSU
  ECE Technical Report}, 2009.

\bibitem{5238753}
M.~Elad and I.~Yavneh, ``A plurality of sparse representations is better than
  the sparsest one alone,'' \emph{Information Theory, IEEE Transactions on},
  vol.~55, no.~10, pp. 4701 -- 4714, Oct. 2009.

\bibitem{4011956}
M.~Elad and M.~Aharon, ``Image denoising via sparse and redundant
  representations over learned dictionaries,'' \emph{Image Processing, IEEE
  Transactions on}, vol.~15, no.~12, pp. 3736 --3745, Dec. 2006.

\bibitem{Candes2012}
E.~J. Cand{\`e}s and M.~A. Davenport, ``How well can we estimate a sparse
  vector?'' \emph{Applied and Computational Harmonic Analysis (In Press)},
  2012.

\bibitem{Ye1953043}
F.~Ye and C.-H. Zhang, ``Rate minimaxity of the lasso and dantzig selector for
  the $\ell_q$ loss in $\ell_r$ balls,'' \emph{J. Mach. Learn. Res.}, vol.~11,
  pp. 3519--3540, Dec. 2010.

\bibitem{6034739}
G.~Raskutti, M.~Wainwright, and B.~Yu, ``Minimax rates of estimation for
  high-dimensional linear regression over $\ell_q$-balls,'' \emph{Information
  Theory, IEEE Transactions on}, vol.~57, no.~10, pp. 6976 -- 6994, Oct. 2011.

\bibitem{Rudelson2009}
M.~Rudelson and R.~Vershynin, ``Smallest singular value of a random rectangular
  matrix,'' \emph{Communications on Pure and Applied Mathematics}, vol.~62,
  no.~12, pp. 1707--1739, 2009.

\bibitem{Tropp20081}
J.~A. Tropp, ``On the conditioning of random subdictionaries,'' \emph{Applied
  and Computational Harmonic Analysis}, vol.~25, no.~1, pp. 1 -- 24, 2008.

\bibitem{Strohmer2003257}
T.~Strohmer and R.~W.~H. Jr., ``Grassmannian frames with applications to coding
  and communication,'' \emph{Applied and Computational Harmonic Analysis},
  vol.~14, no.~3, pp. 257 -- 275, 2003.

\end{thebibliography}

%
%

%

%
%
%




\end{document}